\documentclass{emulateapj}
\usepackage{graphicx}

\begin{document}

\newcommand{\msun}{\mbox{$\:M_{\sun}$}}
\newcommand{\expn}[2]{\mbox{$#1 \times 10^{#2}$}}
\newcommand{\expu}[3]{\mbox{\rm $#1 \times 10^{#2} \rm\:#3$}}
\newcommand{\OIGS}{\:{\rm ergs\:cm^{-2}\:s^{-1}\:\AA^{-1}}}
\newcommand{\LUM}{\:{\rm ergs\:s^{-1}}}
\newcommand{\pow}[2]{\mbox{$\rm10^{#1}\rm\:#2$}}
\shorttitle{SS~Cyg and WX Hyi in Quiescence} \shortauthors{Long et al.}
\title{FUV Spectroscopy of the Dwarf Novae SS~Cygni and WX Hydri\\ in Quiescence}

\author{Knox S.\ Long}
\email{long@stsci.edu}
\affil{Space Telescope Science Institute, 3700 San Martin Drive,
Baltimore, MD 21218}

\author{Cynthia S.\ Froning}
\email{cfroning@casa.colorado.edu}
\affil{Center for Astrophysics and Space Astronomy, \\
University of Colorado, 593 UCB, Boulder, CO, 80309}

\author{Christian Knigge}
\email{christian@astro.soton.ac.uk}
\affil{Department of Physics \& Astronomy,\\
University of Southampton, \\ Southampton SO17 1BJ UK}

\author{William P.\ Blair}
\email{wpb@pha.jhu.edu}
\affil{Department of Physics and Astronomy, Johns Hopkins University,
Baltimore, MD 21218 }

\author{Timothy R.\ Kallman}
\email{tim@xstar.gsfc.nasa.gov}
\affil{NASA/Goddard Space Flight Center, Laboratory for High Energy
  Astrophysics, Code 662, Greenbelt, MD 20771}

\and

\author{Yuan-Kuen Ko}
\email{yko@cfa.harvard.edu}
\affil{Harvard Smithsonian Center for Astrophysics, 60 Garden Street, MS
50, Cambridge, MA 02138}

\begin{abstract}

We present time-resolved FUV spectra of the dwarf novae SS~Cyg and
WX~Hyi in quiescence from observations using the Hopkins
Ultraviolet Telescope on the Astro-1 and Astro-2 Space Shuttle
missions and the Goddard High Resolution Spectrograph on the {\it
Hubble Space Telescope}.  Both dwarf novae are characterized by
blue continua that extend to the Lyman limit punctuated by broad
emission lines including transitions of \ion{O}{6}, \ion{N}{5},
\ion{Si}{4}, and \ion{C}{4}.  The continuum of WX~Hyi can be fit
with a white dwarf model with physically reasonable model
parameters, but neither system actually shows unambiguous
signatures of white dwarf emission. The shape and flux of the
spectrum of SS~Cyg cannot be self-consistently reconciled with a
white dwarf providing all of the FUV continuum flux. Combination
white dwarf/disk or white dwarf/optically thin plasma models
improve the fit but still do not give physically reasonable model
parameters for a quiescent dwarf nova. Assuming that the UV
emission lines arise from the disk, the line shapes indicate that
surface fluxes fall roughly as $R^{-2}$ in both systems. Fits to
the double-peaked line profiles in SS~Cyg indicate that the FUV
line forming region is concentrated closer to the white dwarf than
that of the optical lines and provide no evidence of a hole in the
inner disk. Although the flux from SS~Cyg was relatively constant
during all of our observations, WX~Hyi showed significant
variability during the GHRS observations.  In WX~Hyi, the line and
continuum fluxes are (with the exception of \ion{He}{2}) highly
correlated, indicating a link between the formation mechanisms of
the line and continuum regions.

\end{abstract}

\keywords{accretion, accretion disks --- binaries: close
--- novae, cataclysmic variables --- stars: individual (SS~Cyg) --- stars:
--- individual (WX~Hyi) --- ultraviolet: stars}

\section{Introduction} \label{sec_intro}

Dwarf novae (DN) are close binary systems in which a main-sequence
secondary star loses mass to a non-magnetic white dwarf (WD) via
Roche Lobe overflow onto an accretion disk.  In the standard
accretion theory, the disk radiates away half of the available
energy, while the other half is released in a small boundary layer
(BL) between the inner disk and the WD surface.  DN undergo
quasi-periodic outbursts of 3~--~6 magnitudes due to a thermal
instability in the accretion disk that causes a rapid transition
from a low-temperature, low-accretion rate state to a
high-temperature, high-accretion rate state.  During periods of
quiescence, the mass transfer rate through the disk is low ($\leq
10^{15}$~g~s$^{-1}$) and the disk is cool ($T_{disk} \lesssim
8000$ K) and optically thin in the UV.  On the other hand, the BL
is likely to be extremely hot (T$_{BL} \sim 10^{8}$~K) and
optically thin. As a result, neither the disk nor the BL are
expected to produce large amounts of far-ultraviolet (FUV) flux.
Furthermore, since the companion stars are typically K or M
dwarfs, they are not significant FUV sources. As a result, the
dominant source of FUV emission in quiescent DN is usually assumed
to be the WD. Even if the WD were to have cooled in the interval
between the common envelope phase and the initiation of
mass-exchange, it would have been reheated to temperatures of
10,000 to 50,000~K once mass transfer began\citep{townsley2003}.

As a result of these expectations, FUV spectra of DN in quiescence
have usually been interpreted assuming that the WD dominates, or
at least contributes a significant fraction of, the FUV light from
the system \citep{verbunt1987,deng1994}.  The evidence of
WD-dominance is compelling for a few systems, most notably U~Gem,
VW~Hyi, and WZ~Sge.  In these systems, the WD was first recognized
in {\it IUE\/} spectra from the broad Ly$\alpha$ absorption and the
fact that WD model fits to these spectra yielded temperatures and
luminosities that were consistent with the estimated distances to
these DN \citep{panek1984,mateo1984,holm1988}.  The {\it IUE\/}
results were then confirmed with higher S/N observations made with
the Hopkins Ultraviolet Telescope (HUT), {\it Hubble Space
Telescope\/}  ({\it HST\/}), and the {\it Far Ultraviolet
Spectroscopic Explorer\/} ({\it FUSE\/})
\citep{long1993,long1994a,long2003,sion1994,sion1995,sion2003,froning2001,godon2004}.
In these data, the WDs are recognizable not only from the broad
Ly$\alpha$\ and Ly$\beta$ absorption lines and the overall shape
and flux of the UV spectrum, but also from the relatively narrow
metal absorption lines, which result from the salting of the
atmosphere of the WD by accreting near solar-abundance material.

However, many DN observed in the FUV do not show broad Lyman line
absorption features, nor do they show narrow metal absorption
lines. Instead, numerous DN (including low inclination systems)
show UV spectra characterized by blue continua on which are
superposed broad (typically 10~\AA\ FWHM) emission features
identifiable with resonance lines such as \ion{O}{6}, \ion{N}{5},
\ion{Si}{4}, and \ion{C}{4}.  A variety of explanations for the
emission lines have been advanced, most involving an optically
thin accretion disk \citep{williams1980,tylenda1981,menou2002} or
irradiation of the disk by the boundary layer and the WD
\citep{schwarzenberg1981}. However, the nature of the continuum
emission in quiescent DN has remained controversial. In an attempt
to better characterize possible sources of the FUV continuum and
line emission in quiescent DN, this article describes FUV
observations of two quiescent DN, SS~Cyg and WX~Hyi, obtained with
HUT and with the Goddard High Resolution Spectrograph (GHRS) on
{\it HST\/}.

SS Cyg is an extremely well-studied U-Gem-type DN, with a period,
6.6 hours, that places it well above the period gap.  It is a
fairly low ($\sim$41\degr) inclination system containing a massive
1.2\msun\ WD and a 0.7\msun\ K4 star, which is somewhat
overluminous for a main-sequence star of its spectral type
\citep{friend1990}. SS~Cyg undergoes outbursts that reach a peak
visual magnitude of 8.5, typically lasting either 7 or 15 days and
with an average interoutburst period of 50 days
\citep{cannizzo1992,ak2002}. It is located at a distance of
166$_{-12}^{+14}$ pc, based on astrometry with the {\it HST\/}/FGS
\citep{harrison2000}.  Because it has been observed intensively
since its discovery in 1896, SS~Cyg has been important as one of
the main systems against which detailed predictions of the thermal
instability model for DN outbursts have been measured \cite[see,
e.g.][]{cannizzo1993b}.  The first UV spectra of SS~Cyg were
obtained with {\it IUE\/} and in quiescence show a (F$_\lambda$)
continuum that rises at short wavelengths on which emission lines
from a variety of ionization states are superposed
\citep{fabbiano1981}.  Some, but not all, quiescent {\it IUE\/}
spectra show what appears to be a Ly$\alpha$ absorption feature
that \cite{holm_sscyg1988} interpreted as evidence of emission
from the WD in the system.

WX Hyi, by contrast, is a SU-UMa-type DN, with a period, 1.8
hours, that places it well below the period gap.  It exhibits
normal outbursts that last about 1~day every 11.2 days reaching a
peak magnitude of 12.7, and superoutbursts lasting about 10 days
occurring with a mean interval of 185 days reaching a magnitude of
11.4  \citep{bateson1986}. The inclination is thought to be comparable
to SS Cyg, and the mass of the WD is estimated to be
0.9$\pm$0.3\msun\ \citep{schoembs1981}, but the secondary has not
been detected. The distance to WX~Hyi is poorly known,
\cite{patterson1984} estimating 100~pc based on a correlation
between H$\beta$ equivalent widths and mass accretion rate, and
\cite{warner1987} estimating 265~pc based on a correlation between
period and magnitude during outburst.  In quiescence, {\it IUE\/}
spectra show a flat continuum spectrum, with emission features due
to N~V, Si~IV, C~IV and He~II \citep{hassall1985}. There is no
indication of absorption near Ly$\alpha$.

The HUT observations of SS~Cyg and WX~Hyi in quiescence, which we
describe here, provide the first quiescent spectra of these DN
that extend the simultaneous wavelength coverage from the Lyman
limit to 1840~\AA\ while the GHRS observations provide spectra
with far higher S/N and resolution than was available for these
objects with {\it IUE\/}. The remainder of this paper is organized
as follows: The observations and calibration of the data are
summarized in \S~\ref{sec_obs}. Modeling of the continuum and
emission lines is presented in \S~\ref{sec_analysis}. The results
are discussed in \S~\ref{discussion}, and we present our
conclusions in \S~\ref{conclusions}.

\section{Observations} \label{sec_obs}

\subsection{Hopkins Ultraviolet Telescope} \label{sec_hut}

HUT, which flew on the Space Shuttle as part of the Astro-1 and
Astro-2 missions in 1990 and 1995, utilized a 0.9~m primary mirror
and a prime-focus Rowland circle spectrograph feeding a 1-d photon
counting detector to obtain moderate resolution (R $\sim$3~\AA)
spectra of sources covering the wavelength range 820 -- 1840~\AA\
\citep{davidsen1992,kruk1995}.  SS~Cyg was observed once on
Astro-1 and on Astro-2 through 18 and 20$\arcsec$ diameter
circular apertures, respectively. WX~Hyi was observed three times
on Astro-2 through the 20$\arcsec$ aperture. Observation times are
listed in Table~\ref{tab_obs}.

On Astro-1, the observation of SS~Cyg took place shortly after a ``wide''
$\sim$16 day outburst of the system.  The lightcurve obtained by the
American Association of Variable Star Observers (AAVSO) indicates that
SS~Cyg had in fact fallen to its quiescent level of about 12th magnitude,
but only within a few days of the observation.  It had been below magnitude
11.5 for about 4 days.  The next outburst would not occur for 37 days.  In
contrast, the Astro-2 observation occurred at the end of a cycle, about 20
days after the previous ``narrow'' outburst.  The next outburst would take
place 8~days later.

Because it is relatively faint and in the southern hemisphere,
coverage of the light curve of WX~Hyi is far less complete than
for SS~Cyg.  However, as a result of a special effort by the AAVSO
during the Astro-2 mission, we know that a narrow outburst of
WX~Hyi took place about 1~day prior to the first HUT observation
of WX~Hyi, and that WX~Hyi remained below magnitude 14, and
probably less than magnitude 15, for all three observations of the
system.  Thus, the observations occurred within a single
inter-outburst interval, about 1, 2.5, and 3.1 days after the
outburst.  Normal outbursts in WX~Hyi are typically spaced on 11.2
day intervals \citep{ak2002}. One AAVSO observer reported WX~Hyi
in outburst about 12 days after the last HUT observation, which
would be roughly consistent with our time line, but there are no
confirming observations.  A better observed outburst took place
about 21 days after the last HUT observation.

\begin{figure}
\epsscale{.7}
\plotone{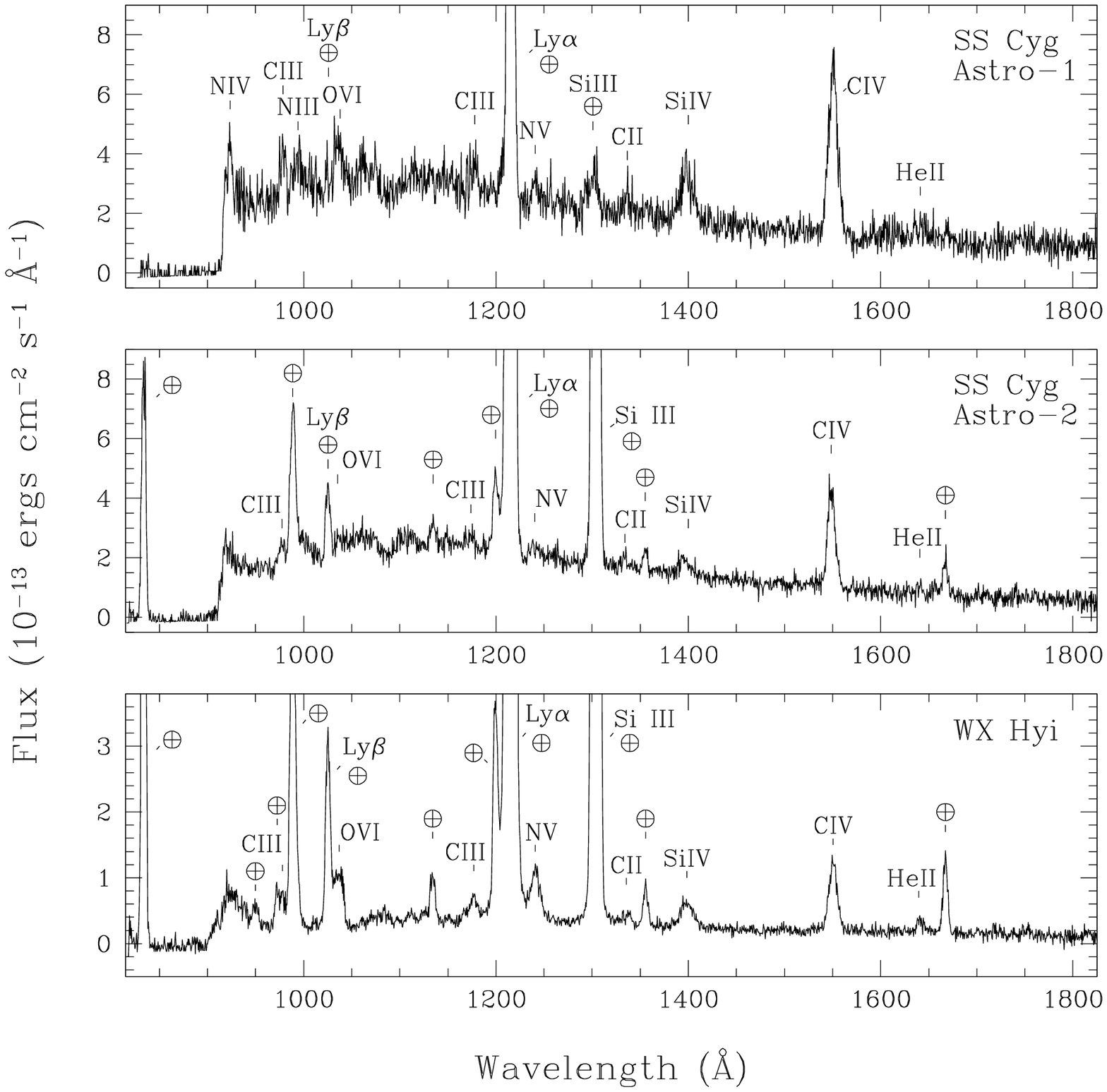}
\caption{The time-averaged HUT spectra of SS~Cyg and
WX~Hyi in quiescence.  The emission lines are labelled and
prominent airglow emission is marked with circled crosses.
\label{fig_hut_spec}}
\end{figure}

For the purposes of this discussion of the HUT observations of
SS~Cyg and WX~Hyi, we have used the final processed data files
delivered to the NSSDC, and available through the Multimission
Archive at Space Telescope. Unless specifically noted, we have
used the photometrically --- but not image-motion
--- corrected data to maximize the counting statistics in the
spectra of these relatively faint sources. Most of the
observations of SS~Cyg and of WX~Hyi took place during the daytime
portion of the shuttle orbit.  However, the Astro-1 observation of
SS~Cyg contained 814~s of data from orbital night. Since airglow
contaminates a significantly greater portion of spectra obtained
with HUT during orbital day than during orbital night, nighttime
data are preferred when available. Therefore, for the Astro-1
observation of SS~Cyg, we have used the 814~s obtained in orbital
night in the analysis that follows. Spectra of the Astro-1 and
Astro-2 observations of SS~Cyg and of the time-averaged (4268~s)
Astro-2 spectrum of WX~Hyi are shown in Figure~\ref{fig_hut_spec}.
The continuum flux levels are close to values reported with {\it
IUE\/} for SS~Cyg and for WX~Hyi in quiescence
\citep{fabbiano1981,ladous1990,hassall1985}.

\subsection{GHRS} \label{sec_ghrs}

The {\it HST\/} observations of SS~Cyg and WX~Hyi were carried out
in the fall of 1996 using the GHRS \citep{heap1995}.  Three
observations of each system were made on approximately one month
centers using the G140L grating and the LSA aperture
(1.74$\arcsec$), yielding a spectral resolution of $\sim$2000. Two
grating settings covering the wavelength ranges 1150 -- 1435~\AA\
and 1377 -- 1663~\AA\ were utilized for each target. For SS~Cyg,
each 3-orbit observation consisted of two segments, one at each
grating setting, which were further divided into individual
exposures of 108~s.  As a result, twenty spectra at each grating
setting were obtained during each of the three observations of
SS~Cyg.  A similar procedure was used for WX~Hyi. However, since
WX~Hyi was observed in {\it HST\/}'s continuous viewing zone, each
observation was divided into four segments, resulting in forty
(108 s integration) spectra  at each grating setting.

To ease scheduling (especially for WX~Hyi, which has a short inter-outburst
period), the observations were not scheduled as targets of opportunity, and
as it turned out, two of the observations of each system occurred during an
outburst.  We will confine ourselves here to the low-state data, obtained
on 26 September for SS~Cyg and on 5~August for WX~Hyi (see
Table~\ref{tab_obs}).  AAVSO lightcurves indicated that the 26 September
observation of SS~Cyg occurred in mid-quiescence, about 17 days after
return to optical quiescence from a ``wide'' outburst of SS~Cyg, and about
12 days before a ``narrow'' outburst.  The sparse AAVSO data on WX~Hyi show
that the 5~August observation occurred near the end of a quiescent
interval, about 8~days after a normal outburst, and 1~day before another
normal outburst.

\begin{figure}
\epsscale{.65}
\plotone{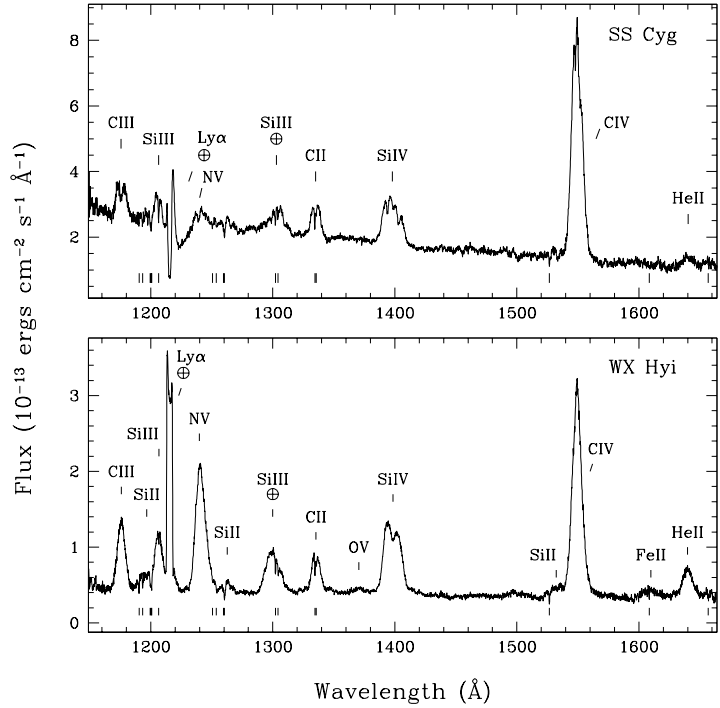}
\caption{The time-averaged GHRS spectra of SS~Cyg and
WX~Hyi in quiescence.  The emission lines are labelled.  The
vertical lines below each spectrum indicate the location of the
interstellar absorption features.  Emission lines that are or may
be affected by airglow are labelled with circled crosses.
\label{fig_ghrs_spec}}
\end{figure}

For the analysis of the {\it HST\/} spectra described here, we
recalibrated the observations using the standard (CALHRS) pipeline
with calibration files available in 1999 February, which properly
account for the declining sensitivity of the GHRS at the shortest
wavelengths.  For wavelength calibration, we elected to register
using interstellar lines, although in fact the difference between
this and the internal lamp calibration was small ($<$0.1~\AA).
Time-averaged spectra of the low-state GHRS observations are shown
in Figure~\ref{fig_ghrs_spec}.

\section{Analysis} \label{sec_analysis}

The FUV spectra of SS~Cyg and WX~Hyi are characterized by blue
continua on which are overlaid strong, broad emission lines from
resonance and excited state transitions of ionized metals and
\ion{He}{2}.  SS~Cyg also has a broad Ly$\alpha$ absorption
feature, which is not seen in WX~Hyi. WX~Hyi has a richer line
spectrum than SS~Cyg, with the former showing emission from
\ion{O}{6}, \ion{N}{5}, and \ion{He}{2}, all of which are weak or
absent in the latter. The line profiles are double-peaked in
SS~Cyg, consistent with emission from a Keplerian disk, but are
single-peaked in WX~Hyi.

There were no dramatic long time scale variations in the shapes of
the lines or continuum over the baseline of the HUT observations
for either object.  The three HUT observations of WX~Hyi were
obtained during a single quiescent cycle.  Between the first and
second observations, the continuum declined by 50\% but then
rebounded in the third observation to a flux midway between the
first two (e.g., the 1500~\AA\ continuum fluxes were
\expn{3.3}{-14}, \expn{1.7}{-14}, and \expu{2.3}{-14}{\OIGS}). The
behavior of the continuum is inconsistent with the report of a
steady, long-term decline in the UV line and continuum fluxes
given by \citet{hassall1985}.  It is also inconsistent with the
predictions of the standard disk instability model for DN, which
predicts an increase in the disk flux during quiescence
\citep{lasota2001}. In SS~Cyg, the continuum level also varied
little between the two quiescent observation epochs (spanning a
range of six years).  The 1500~\AA\ flux was 30\% brighter in the
first observation and the emission line strengths in \ion{Si}{4}
$\lambda\lambda$1398,1402 and \ion{C}{4} $\lambda$1550 were
$\sim$60\% brighter.

Below, we discuss model fits to the continuum, emission line profiles, and
short time scale variability in the FUV spectra of SS~Cyg and WX~Hyi.

\newpage
\subsection{Continuum emission} \label{sec_cont}

As noted earlier, the traditional interpretation of FUV continua
in quiescent DN has involved emission from WDs with surface
temperatures of 10,000 -- 50,000~K. For SS~Cyg, estimates of the
temperature of the WD have ranged from 34,000 -- 40,000~K based
upon the slope of the continuum and the existence of Ly$\alpha$\
absorption in some of the {\it IUE\/} spectra
\citep{fabbiano1981,holm_sscyg1988}.  For WX~Hyi, there are no
previous estimates of the WD temperature.

In an attempt to see to what degree a WD explanation for the
continuum emission in SS~Cyg and WX~Hyi is viable, we have
synthesized sets of log g~=~8 WD model spectra using Hubeny's TLUSTY
and SYNSPEC programs \citep{hubeny1988,hubeny1994} convolved to
the resolution of our HUT and GHRS observations. For comparison,
we also synthesized a set of optically-thick steady-state
accretion disk spectra from an appropriately weighted set of
stellar atmospheres, also constructed with TLUSTY and SYNSPEC,
using the procedure described in detail by \citet{long1994b}. In
generating the models we have assumed a WD mass of 1.2~\msun\ and
an inclination of 41$^{\circ}$ for SS~Cyg \citep{friend1990},
while for WX~Hyi we have assumed 0.9~\msun\ and 40$^{\circ}$
\citep{schoembs1981}.  Finally, we have used Cloudy
\citep{ferland1996} to create a set of model spectra of plasmas in
coronal equilibrium. For the thin-plasma spectra we have limited
emission to that from H and He since we are here concerned with
continuum fits to the data.

\begin{figure}
\centerline{\includegraphics[height=.8\textwidth,angle=-90]{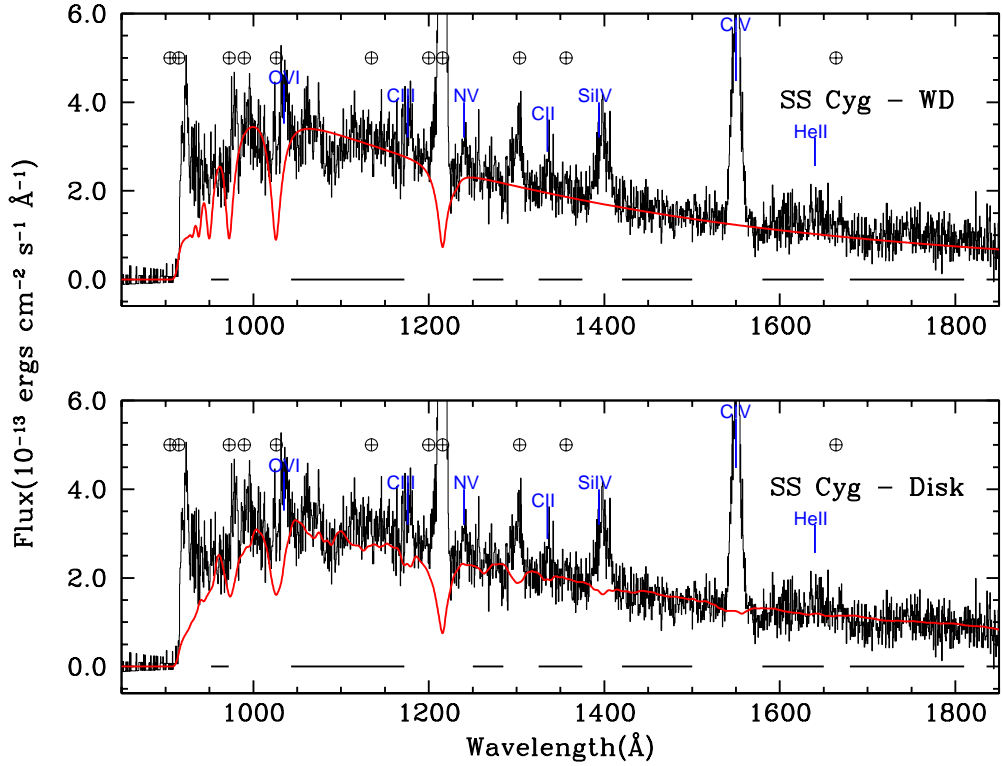}}
\caption{Best-fitting DA WD and accretion disk
model fits to the HUT spectrum of SS~Cyg. The reddening is set to E(B-V)~=
0.04. The regions included in the fit are indicated by the bars under the
spectra. Regions with strong source or airglow lines were
excluded. \label{fig_hut_cyg_cont}}
\end{figure}

\begin{figure}
\centerline{\includegraphics[height=.8\textwidth,angle=-90]{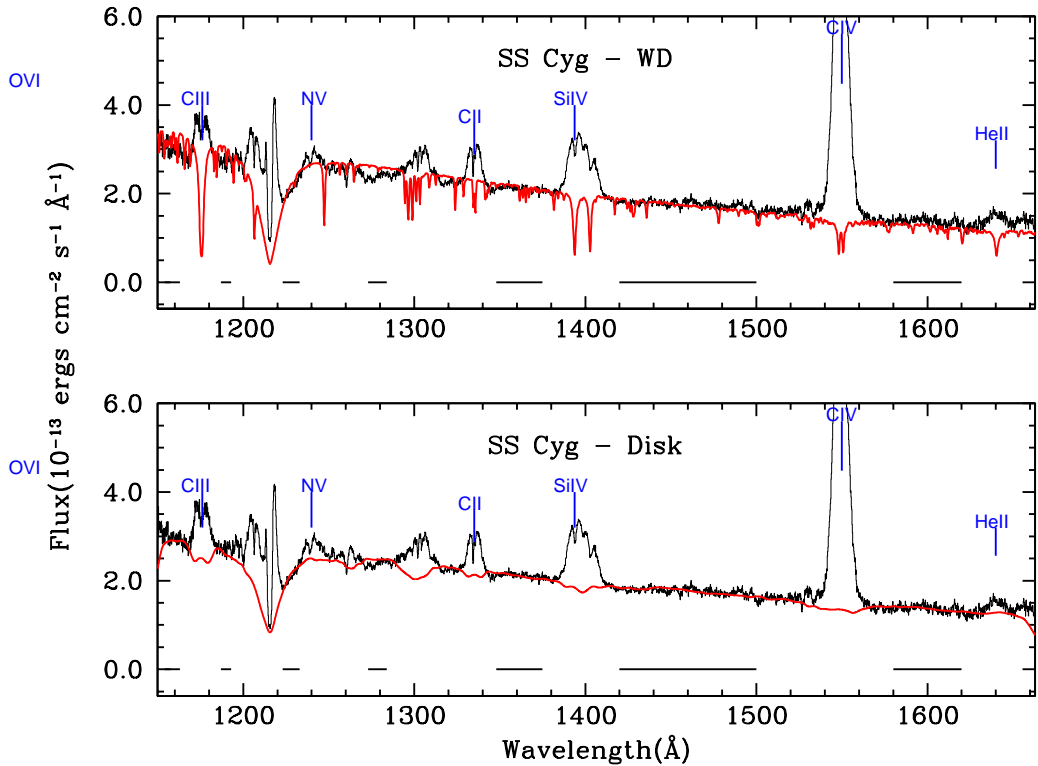}}
\caption{Best-fitting normal abundance WD and
accretion disk model fits to the GHRS spectrum of SS~Cyg.  The
reddening is set to E(B-V)~= 0.04.  \label{fig_ghrs_cyg_cont}}
\end{figure}

\begin{figure}
\centerline{\includegraphics[height=.8\textwidth,angle=-90]{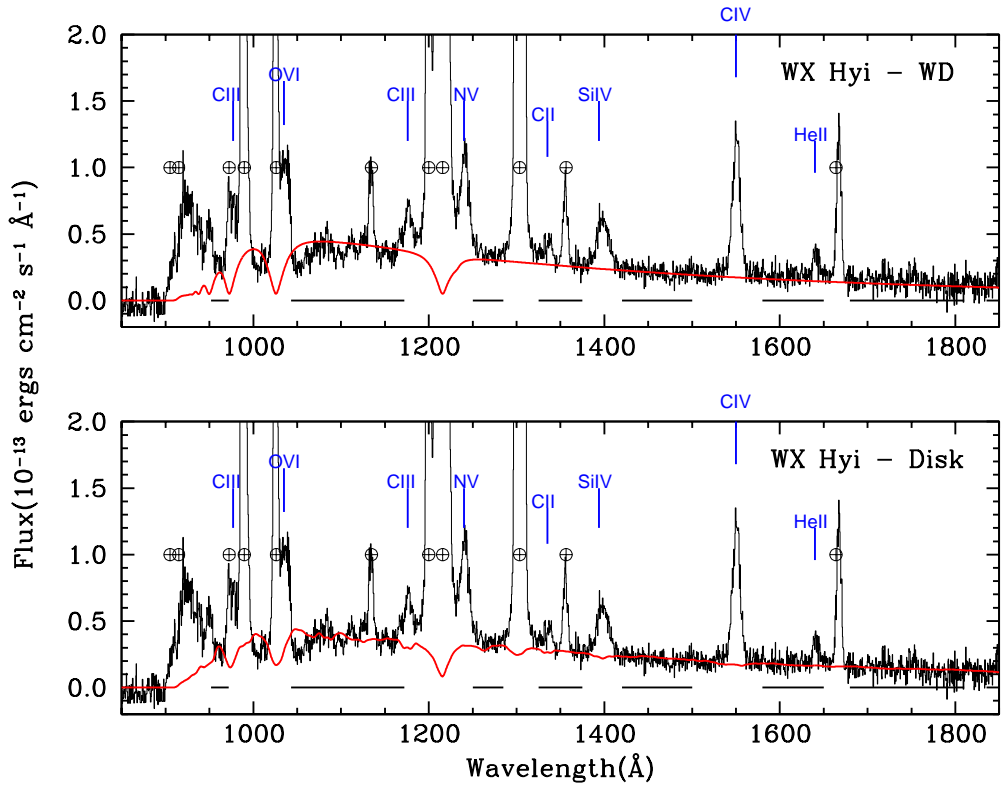}}
\caption{Best-fitting DA WD and accretion disk
model fits to the HUT spectrum of WX~Hyi.  \label{fig_hut_hyi_cont}}
\end{figure}

\begin{figure}
\centerline{\includegraphics[height=.8\textwidth,angle=-90]{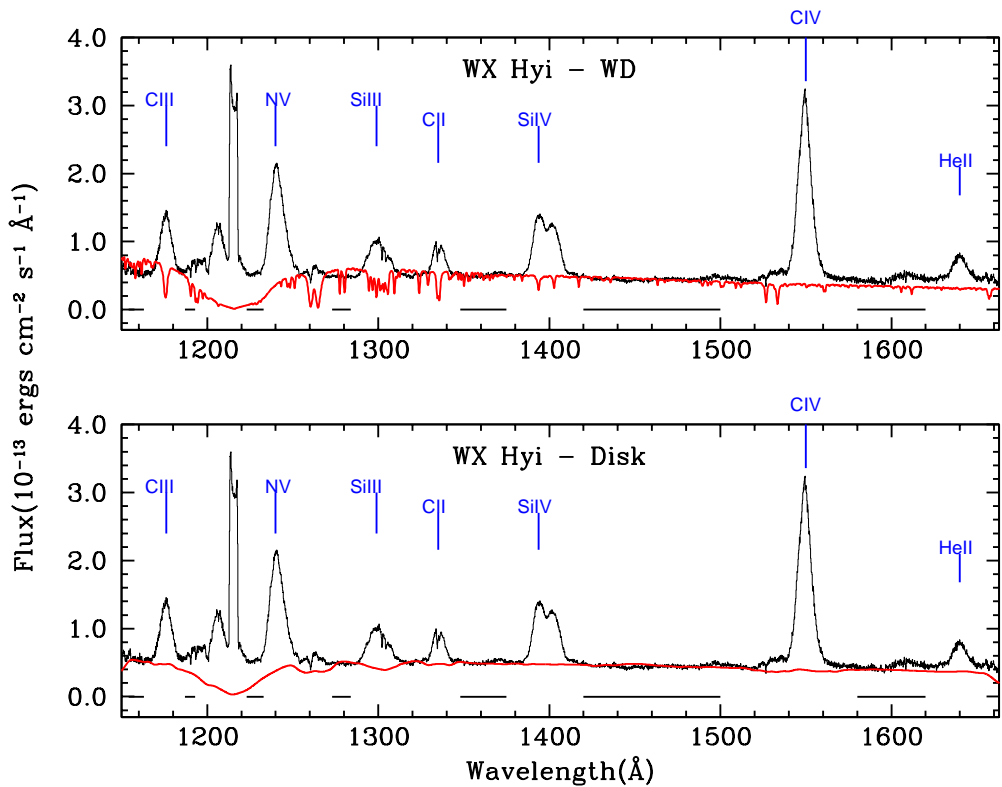}}
\caption{Best-fitting normal abundance WD and
accretion disk model fits to the GHRS spectrum of WX~Hyi.
\label{fig_ghrs_hyi_cont}}
\end{figure}

We first fit spectra assuming plausible values of the reddening
for both SS~Cyg and WX~Hyi: E(B-V)~= 0.00, 0.04, and 0.07. We began
with models consisting of either a simple WD, an optically thick
accretion disk, or a thin plasma.  For WDs, we considered both
pure DA atmospheres and atmospheres with a normal heavy element
composition.  We then considered plausible combinations of WDs and
optically thick disks and WDs and thin plasmas.  Results of the
fits with E(B-V) set to be 0.04 are recorded in
Tables~\ref{sscyg_fits} and~\ref{wxhyi_fits} for SS~Cyg and
WX~Hyi, respectively.   All of the fits to the data were carried
out using standard $\chi^2$ minimization techniques.  Since the
spectra contain emission lines from the disk and (for HUT) from
strong airglow lines, we fit only the ``line-free'' regions of the
spectra. Results for the DA model and for the disk fits for the
SS~Cyg HUT and GHRS data are shown in
Figures~\ref{fig_hut_cyg_cont} and~\ref{fig_ghrs_cyg_cont}, while
Figures~\ref{fig_hut_hyi_cont} and~\ref{fig_ghrs_hyi_cont} show
the fits to the WX~Hyi spectra. Each figure shows the best WD fit
and the best disk fit to the data (for SS~Cyg, the night-only data
from the Astro-1 HUT observations are shown).

For the HUT observation of SS~Cyg, the best fitting DA atmosphere
(assuming E(B-V)~= 0.04) is shown in Figure~\ref{fig_hut_cyg_cont}.
In SS~Cyg, the best fit has a temperature of $\sim$46,000~K and
normalization that corresponds to a WD radius of \expu{8.2}{8}{cm}
at the known distance of 166$_{-12}^{+14}$~pc
\citep{harrison2000}. The model fit is actually reasonable in a
statistical sense ($\chi^2_{\nu} \sim 1.3$) for the fitted
regions, but it is important to note that because of the line
emission we have excluded most, if not all, of the regions that
would contain the strongest signatures of the WD. Best-fitting DA
models with E(B-V)~= 0.0 and 0.07 yield temperatures of 38,000 and
58,000~K, respectively.  Increasing E(B-V) increases the
temperature because the model fits almost entirely reflect the
overall slope of the spectrum.  The temperature derived from
normal-abundance WD models is very similar to that of the DA
model.

For the GHRS observation of SS~Cyg, the best fitting normal
abundance model, again assuming E(B-V)~= 0.04, is shown in
Figure~\ref{fig_ghrs_cyg_cont}. The best DA fit has a somewhat
lower temperature, 40,800~K, than the HUT model fits and a
normalization that corresponds to a WD radius of
\expu{9.7}{8}{cm}. In this case, reflecting the higher statistical
quality of the data, $\chi^{2}_{\nu} \sim 3$. The normal abundance
model fits provide qualitatively and quantitatively poorer
approximations to the GHRS data.  By our choice of wavelength
ranges to fit, we have actually excluded the strongest absorption
lines in the atmosphere of a WD with a temperature of about
37,000~K.  Nevertheless, in the high quality data obtained with
the GHRS, there are many other lines present in the models that
one would expect to observe that are absent in the GHRS spectrum.

If the mass of the WD in SS~Cyg is 1.2~\msun, and if the WD
follows a normal WD mass-radius relationship \citep{panei2000},
then the expected radius of the WD is \expu{\sim4.2}{8}{cm},
considerably smaller than the values we have derived. Assuming the
lower limit on current estimates of the WD mass in SS~Cyg,
1.0~\msun\ \citep{martinez1994}, gives a maximum WD radius
\expu{\sim6.3}{8}{cm}, still well below our derived radii. For a
WD radius in the range suggested by the mass-radius relationship,
the WD temperature (all else remaining equal) would need to be
$\sim$55,000 -- 85,000~K to match the observed continuum flux.
However, model WD spectra at these temperatures are too blue in
the FUV ($<1100$~\AA) to match the observed spectra. Thus it is
unlikely that the WD dominates the FUV spectrum of SS~Cyg.

Disk model fits to the SS~Cyg spectra are also shown in
Figures~\ref{fig_hut_cyg_cont} and~\ref{fig_ghrs_cyg_cont}. Like the WD
models, the disk models approximate the overall slope of the spectrum
fairly well and plausibly reproduce the Ly$\alpha$\ feature. Despite the
fact that the models were constructed with normal abundance atmospheres,
the disk models produce no narrow lines because of Doppler broadening.  For
the HUT and GHRS spectra, the disk models suggest a mass accretion rates of
\expu{1.0}{17}{g\:s^{-1}} (\expu{1.6}{-9}{\msun\ yr^{-1}}) and
\expu{4.3}{16}{g\:s^{-1}} (\expu{6.8}{-10}{\msun\ yr^{-1}}), respectively.
This is larger than the rates expected from a DN in quiescence.  The disk
models are normalized to a distance of 100~pc, and so the expected value of
the normalization for a distance of 166~pc is 0.36, larger than the best
fit normalization would suggest, especially for the higher value of the
accretion rate.

Simple thin plasma model fits to the data are poor both for the
HUT and GHRS spectra of SS~Cyg.  This is because simple thin
models are not sufficiently blue in the wavelength range longward
of 1200~\AA, nor do they turn over rapidly enough to match the
spectrum in the sub-Ly$\alpha$\ range.  The best fits tend to the
highest temperatures in our grid 100,000~K and have emission
measures of order \expu{3}{57}{cm^{-3}}.\footnote{We note in
passing, that these parameters we derive are not consistent with
the expected properties of the BL or corona in quiescent CVs,
which in models are much hotter, X-ray sources with
T$\simeq10^{8}$~K \citep{narayan1993,meyer1994}.  The problem is
not that the temperature is too low since the spectral shape of a
thin plasmas does not change significantly in the FUV at
temperatures higher than 100,000~K.  Rather, the problem is that if
the temperature were $10^8$~K, then the required emission measure
would be of order \pow{59}{cm^{-3}}, as compared to estimates, in
the case of SS~Cyg, of \expu{2}{56}{cm^{-3}} \citep{done1997}. To
connect this plasma to the BL or high T~corona, one would need to
invoke a more complicated models with a a much tighter connection
to a physical picture of the source of the emission.} Both
combination models, a WD plus a disk or a WD plus a thin plasma,
produce improvements in the model fits to SS~Cyg in most cases. In
particular, the WD plus thin plasma yields the lowest value of
$\chi^2_{\nu} \sim 1.1$ for the HUT data and $\sim1.4$ for the
GHRS.  However, the results remain disappointing in the sense that
they do not really produce physically believable models for
SS~Cyg.

The situation for WX~Hyi is similar, although there is more
variation between the best fits for the HUT and GHRS spectra. For
the HUT spectra of WX~Hyi, the best fitting DA model WD
temperature is 53,800~K and the normalization suggests a radius of
\expu{7.4}{8}{cm} at a fiducial distance of 300~pc. For the GHRS
spectrum and normal abundances, the temperature is lower,
25,000~K, and the radius is larger, \expu{1.9}{9}{cm}.  These are
plausible values (the M-R relation for a 0.9~M$_{\odot}$ WD gives
a radius $\sim$ \expu{7}{8}{cm}), if one assumes that the real
signatures of a metal-enriched WD surface have been obscured by
the disk emission.  As with SS~Cyg, however, there are absorption
features in regions away from the emission lines that appear in
the models but not in the observed spectrum. For disk models and
the HUT spectra, the slope of the spectrum requires a mass
accretion rate of \expu{5.1}{16}{g\:s^{-1}}. However, if this were
actually an accurate description of WX Hyi, then the observed flux
should have been an order of magnitude too brighter than observed,
assuming a distance of 300~pc. On the other hand, for disk models
and the GHRS spectra, one obtains a mass accretion rate of
\expu{3.3}{15}{g\:s^{-1}}, and in this case, the flux expected
from such a disk is only 25\% of that expected if the distance is
300~pc.

So what should one conclude in light of these results?  First,
while it is true that SS~Cyg shows evidence of Ly$\alpha$\
absorption and both SS~Cyg and WX~Hyi have spectral shapes that
have the general curvature expected of a WD, it is dangerous to
assume the spectrum is actually dominated by the WD; a disk model
shows similar features. Second, to make serious progress on
interpreting the spectra of DN in quiescence it is going to be
necessary to develop better models of other components in the
system.  We will return to this topic in \S~\ref{discussion}.

\subsection{Time-averaged emission spectra}

\subsubsection{Modeling the emission line profiles} \label{sec_profiles}

If we assume that the emission lines observed in SS~Cyg and WX~Hyi
are produced on a surface layer of the disk, then the shape of the
emission lines constrain the region of the disk where the emission
lines originate. This is because, in  a geometrically thin, flat
accretion disk in Keplerian rotation, Doppler broadening dominates
the line broadening for optically thin lines. By assuming a local
line surface brightness profile, such as $f(R) \propto
R^{-\alpha}$, and summing over the contribution of the entire
disk, a model emission line profile can be made that depends on
the radial extent of the line-forming region (or alternatively,
the ratio of the inner to outer disk radii), the velocity at the
disk edge, and the value of the power law index, $\alpha$
\citep{smak1981}.  Such models produce the classic double-peaked
emission line profiles characteristic of accretion disks in
Keplerian rotation. In these models, $\alpha$~sets the shape of
the line wings, the radial extent of the disk controls the extent
of the line wings, and the velocity at the disk edge determines
the separation of the line peaks.  For optically thick emission
lines, shear broadening becomes important, and the line profiles
depend additionally on inclination \citep{horne1986,orosz1994}.
Including shear for optically thick lines has the primary effect
of deepening the central minimum between the line peaks at
moderate to high binary inclinations.

\begin{figure}
\plotone{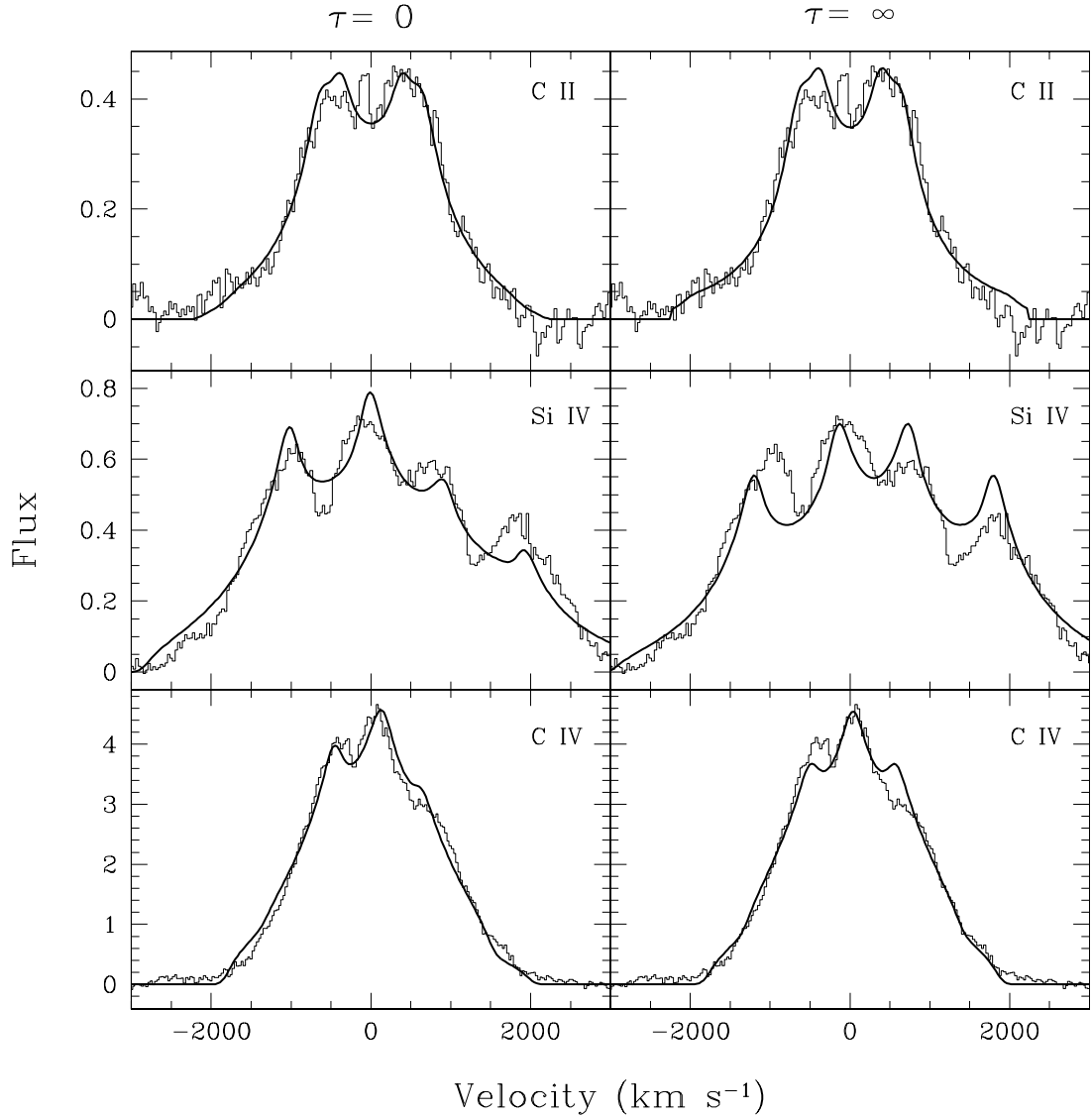}
\caption{Model fits to the emission line profiles in
SS~Cyg.  The continuum has been normalized to 1 and subtracted from the
line profiles.  From top to bottom, \ion{C}{2}, \ion{Si}{4}, and
\ion{C}{4}\ are shown.  Fits in the left panels assume optically thin
emission lines, while fits in the right panel are for optically thick
lines. \ion{N}{5}\ and \ion{He}{2}\ are not fit due to the weakness of
these lines. \label{fig_sscyg_profiles}}
\end{figure}

\begin{figure}
\plotone{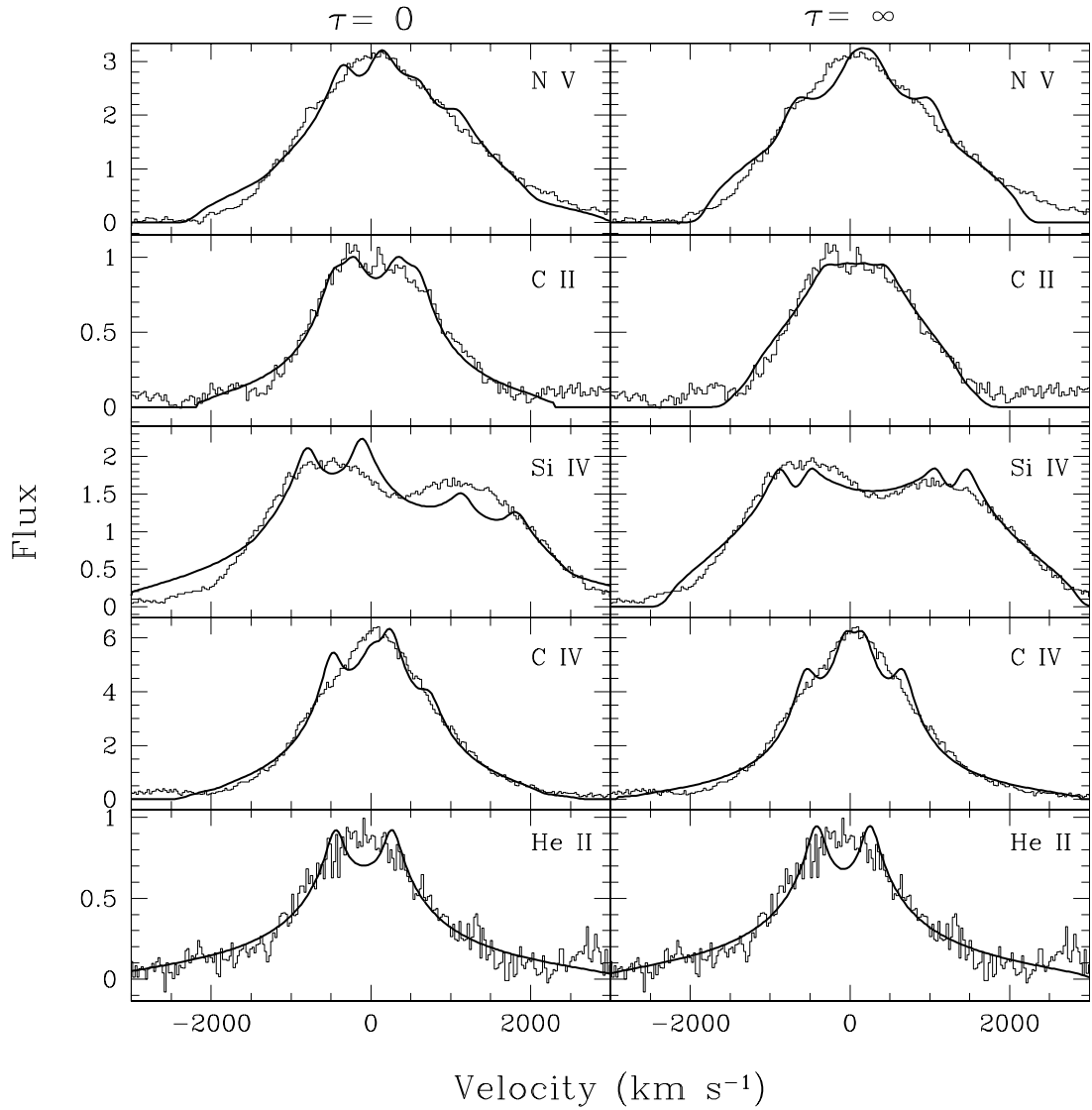}
\caption{Model fits to the normalized and continuum
subtracted emission line profiles in WX~Hyi.  \ion{N}{5}, \ion{C}{2},
\ion{Si}{4}, \ion{C}{4}, and \ion{He}{2}\ are shown from top to bottom.
The models in the left panel are for optically thin lines, while the models
in the right panel are for optically thick
lines. \label{fig_wxhyi_profiles}}
\end{figure}

Following the formalism of \citet{orosz1994}, we modeled the line
profiles of the strong UV emission lines in the averaged GHRS
spectra of SS~Cyg and WX~Hyi.  The doublet nature of most of the
lines and the finite instrument resolution were taken into account
in the fits.  In Figures~\ref{fig_sscyg_profiles} and
\ref{fig_wxhyi_profiles} we compare the shapes of the emission
lines to the best-fitting theoretical line profiles for both the
optically thin and optically thick cases.  The corresponding model
output parameters are given in Table~\ref{tab_profiles}.  For
SS~Cyg, the optically thin line profiles provide better fits to
the observed emission lines.  The deepening of the central
depression between the line peaks for the optically thick models
is a small but appreciable effect at the moderate inclination of
SS~Cyg and worsens the fit to the lines.  The primary evidence for
optically thin UV line emission in SS~Cyg comes from the doublet
ratios of the \ion{Si}{4}\ and \ion{C}{4}\ lines, where the model
fits are clearly better for a ratio of 2:1 (optically thin) than
1:1 (optically thick). The opposite result holds true for WX~Hyi,
in which the optically thick models are more consistent with the
observed line shapes than the optically thin ones. This conclusion
is based on how the models fit the doublet line ratios rather than
how they fit the depths of the line centers relative to the peaks,
since in WX~Hyi the observed line profiles are not double-peaked.
For both SS~Cyg and WX~Hyi, the power law index, $\alpha$, varies
from 1.8 -- 2.5 for the different emission lines (with a few large
excursions from this range, such as the optically thick model of
\ion{N}{5} in WX~Hyi). The value of $\alpha$ controls the shape of
the line wings in the models, and it is clear from
Figures~\ref{fig_sscyg_profiles} and~\ref{fig_wxhyi_profiles} that
our models are not always descriptive of the observed line wings.
In general, however, the surface brightness varies roughly as
$f(R) \propto R^{-2}$ for the UV emission lines in SS~Cyg and
WX~Hyi, which is consistent with the surface brightness
distributions of optical emission lines in CVs \citep[see][and
sources therein]{robinson1993}.

In SS~Cyg, the observed emission lines are double-peaked.  For a
Keplerian accretion disk, the velocity separation between the
peaks corresponds to twice V$_{K} \sin \imath$ at the outer radius
of the line formation region, giving a lower limit to the size of
the disk.  The best-fit optically thin line profile models for
SS~Cyg give V$_{K} \sin \imath$ equal to 490$^{+75}_{-33}$,
470$^{+52}_{-44}$, and 280$^{+19}_{-60}$ km~s$^{-1}$ for
\ion{C}{2}, \ion{Si}{4}, and \ion{C}{4}, respectively.  For the
Balmer series, \citet{martinez1994} give V$_{K} \sin \imath$~=
178, 245, 294, and 307~km~s$^{-1}$ for H$\alpha$ to H$\delta$.  If
all of the emission lines originate in a fully Keplerian accretion
disk, the velocity separations of the UV line peaks in SS~Cyg
indicate that their line formation regions cut off at smaller
radii (higher disk velocity) than those of the optical emission
lines.  The strongest UV line, \ion{C}{4}, does extend to outer
disk radii comparable to those of the Balmer emission lines above
H$\alpha$.

\subsubsection{Searching for a hole in the inner accretion disk}

Although DN outbursts are in most respects well described by the
thermal-viscous disk instability model \citep[see, e.g.,][for a
review]{cannizzo1993}, the standard model is unable to account for
the 0.5--1~d delay between the start of outbursts at optical and
UV wavelengths observed in several DN \citep[e.g.,][]{hassall1983,
polidan1984, verbunt1987}. A number of theoretical models have
attempted to explain the optical--UV lag by arguing that the inner
accretion disk is disrupted during the inter-outburst interval and
must be replenished before the UV outburst can begin. Mechanisms
for disrupting the inner disk include a coronal siphon
\citep{meyer1994}, a weak WD magnetic field \citep{livio1992}, or
irradiation by the hot WD \citep{king1997}.  Typical models
predict that the inner radius of the accretion disk moves out to
$R_{inner}$ \expu{\sim3-4.5}{9}{cm} far from outburst
\citep{liu1997,king1997}.

Such a hole in the accretion disk should be observable in high excitation
emission lines (i.e., those we most expect to be associated with the inner
disk), and would be reflected in a loss of the high-velocity line wings.
We undertook to determine the inner radii of the UV line emitting regions
in SS~Cyg (for which the double-peaked emission lines strongly support an
accretion disk origin).  Assuming Keplerian disk rotation, the inner radius
of the line formation region is related to the observed maximum velocity of
the emission line (or the HWZI) by
\begin{equation}
v_{max} = \sqrt{ \frac{G M_{WD}}{R_{inner}} } \sin i
\end{equation}
where we set M$_{WD}$ = 1.2~M$_{\odot}$ and $i$ = 41$^{\circ}$. We
determined $v_{max}$ from the model fits to the normalized
emission line profiles (Table~\ref{tab_profiles}). We used the
optically thin models for SS~Cyg, although the resulting values of
R$_{inner}$ do not depend on details of the model fitting but only
on a correct measure of the maximum velocity in the line wings. We
found R$_{inner}$ = 1.7$^{+0.3}_{-1.4} \times$10$^{9}$,
1.3$^{+0.7}_{-0.8} \times$10$^{9}$, and 2.4$^{+2.4}_{-1.0}
\times$10$^{9}$~cm for the \ion{C}{2}, \ion{Si}{4}, and \ion{C}{4}
emitting regions.  As a check, we also directly estimated the HWZI
of the emission lines using Gaussian fits to the doublets; our
results were consistent with those quoted above to within the
error bars.

Thus, the inner radii of the UV emission line formation regions
are 2~to~4 times smaller than the size of the hole in the disk
predicted by theoretical models.  If the UV emission lines in
SS~Cyg originate in a Keplerian accretion disk, there is no
evidence for substantial disruption in the inner accretion disk in
quiescence. The complex emission line profiles observed in SS~Cyg
and the lack of double-peaked profiles in WX~Hyi and other CVs
underscores that our current emission line models are incomplete,
however; sub-Keplerian rotation and/or the presence of vertically
extended line emission could still allow for a hole in the inner
accretion disk during quiescence.
\bigskip

\subsection{Variability in the emission lines and the continuum} \label{sec_var}

As noted earlier, the GHRS observations comprise a series of
spectra with individual integration times of 108~s.  Given that
the spectra are individually of high S/N, we investigated whether
the spectra exhibit variations on this time scale, since time
variations might provide clues to the nature of the emitting
regions in the lines or the continuum, or both.

\begin{figure}
\plotone{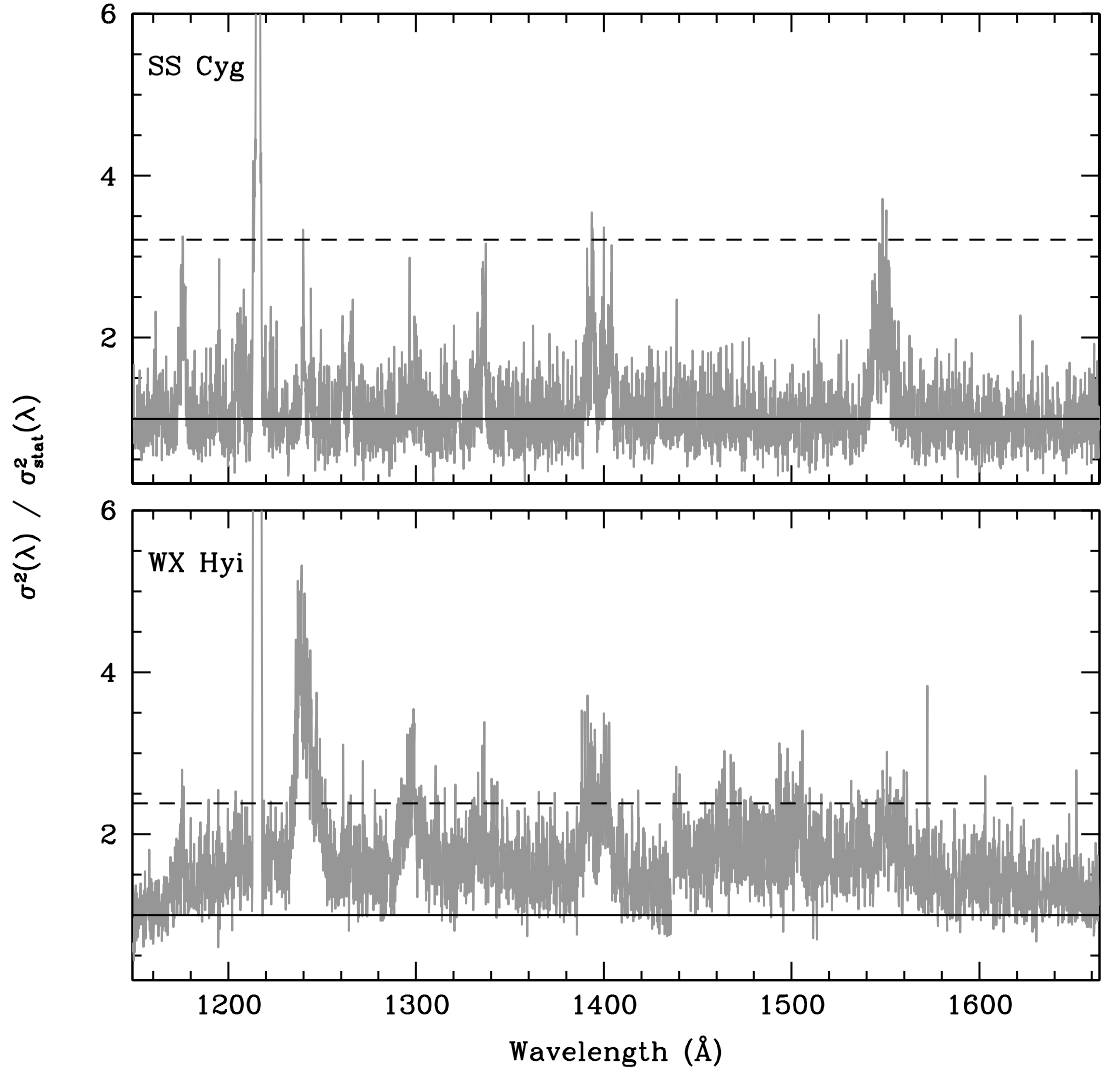}
\caption{The ratio of the observed variance in the spectra
of SS~Cyg and WX~Hyi to the variance expected from counting statistics.
The solid line indicates a ratio of~1, the expectation value for no
intrinsic source variability.  The dashed line shows the 1\% probability
line for a single point exceeding this level in the case of no intrinsic
variability. \label{fig_var}}
\end{figure}

From our initial inspection of the data, it was clear that there
were large difference in the individual spectra of WX~Hyi, but
that, to first order, SS~Cyg was not varying. To begin to
characterize the variability, or lack thereof, in the two
datasets, we first computed the fractional variance spectrum,
\begin{equation}
\sigma^2(\lambda)
= \frac{1}{N-1} \Sigma_{n=1,N} \left(\frac{f_n(\lambda)}{<f(\lambda)>}-1\right)^2
\end{equation}
where N is the total number of spectra, $f_{n}(\lambda)$ is the
observed flux in the n$^{th}$ spectrum, and $<f(\lambda)>$ is the flux
in the mean spectrum. We have then compared this to the variance
expected due to counting statistics, i.e.,
\begin{equation}
\sigma_{stat}^2(\lambda) = \frac{1}{N} \Sigma_{n=1,N} \left(\frac{E(f_n(\lambda))}{<f(\lambda)>}\right)^2
\end{equation}
where $E(f_n(\lambda))$ is the uncertainty in the n$^{th}$ spectrum.
Results of this evaluation are presented in Figure~\ref{fig_var}, which
shows the fractional variance spectrum of each object normalized with
respect to the expected statistical variance.  The solid and dashed lines
indicate, respectively, the expectation value and the value at which there
is a 1\% probability of a single point exceeding that level for no
intrinsic variability \citep{knigge1997}.

This analysis confirmed the impressions obtained from our
inspection of the individual spectra.  For SS~Cyg, there is no
evidence of variability in the continuum; the ratio of the
fractional variance to the variance expected from counting
statistics is close to~1 at all wavelengths. Several of the
emission lines show low-level variability.  The strongest
non-statistical variation is seen in Ly$\alpha$, but this feature
is contaminated by terrestrial airglow. \ion{C}{4} shows the
strongest variability of the non-contaminated emission lines.
Examination of fluxes in the individual spectra places the
1-$\sigma$ continuum variability at $<$4\%, while the peak
\ion{C}{4} line flux has a maximum deviation from the mean flux of 12\%.

In WX~Hyi, on the other hand, both the continuum and the emission
lines show variability in excess of purely statistical variation
at all wavelengths.  Examination of spectra shows that variations
occurred on time scales at least as short as individual exposures.
The strongest source variation is seen in \ion{N}{5}. In order to
determine the extent to which the emission line and continuum
variations are correlated in WX~Hyi, we fit the strong emission
lines and the underlying continua in the individual spectra of
each system using SPECFIT \citep{kriss1994}. We fit each line
locally with a flat continuum and a Gaussian emission profile.  We
modeled the doublet lines with two Gaussians with equal FWHM and
the relative positions of the doublet transitions fixed.  The
other emission features were modeled as single lines. We verified
that the Gaussian fits agreed to within the uncertainties
determined by SPECFIT to a direct calculation of the line EW
obtained by summing over the line. Deviations between the two
methods almost always depended on the placement of the continuum
level, which the statistical fit of SPECFIT is best equipped to determine.

\begin{figure}
\plotone{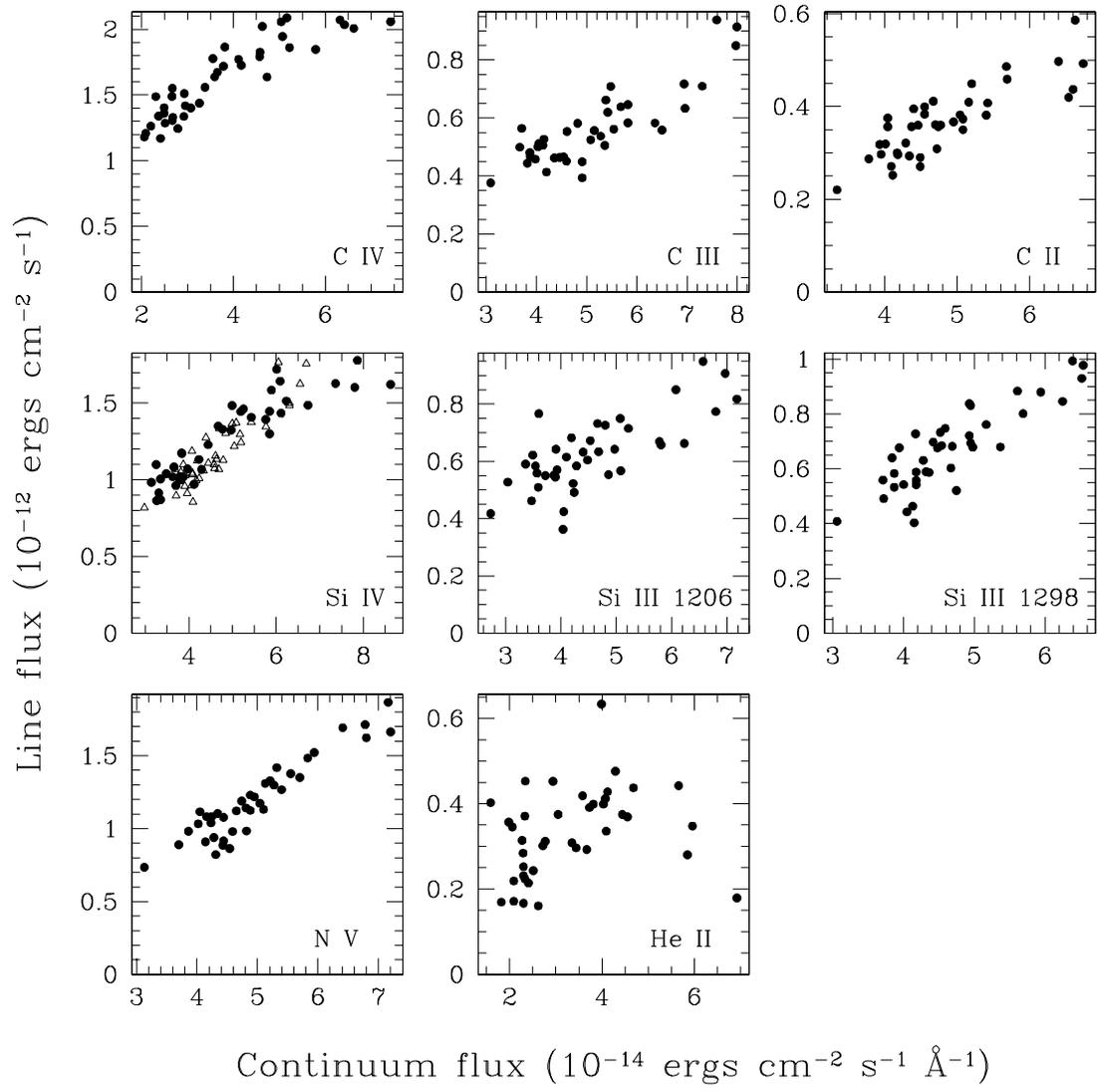}
\caption{Emission line flux versus continuum flux level
for the strong emission lines in WX~Hyi.  For \ion{Si}{4}, the open
triangles show frames acquired in the short wavelength grating setting,
while the filled circles were taken in the long wavelength
setting. \label{fig_wxhyi_flux}}
\end{figure}

\begin{figure}
\plotone{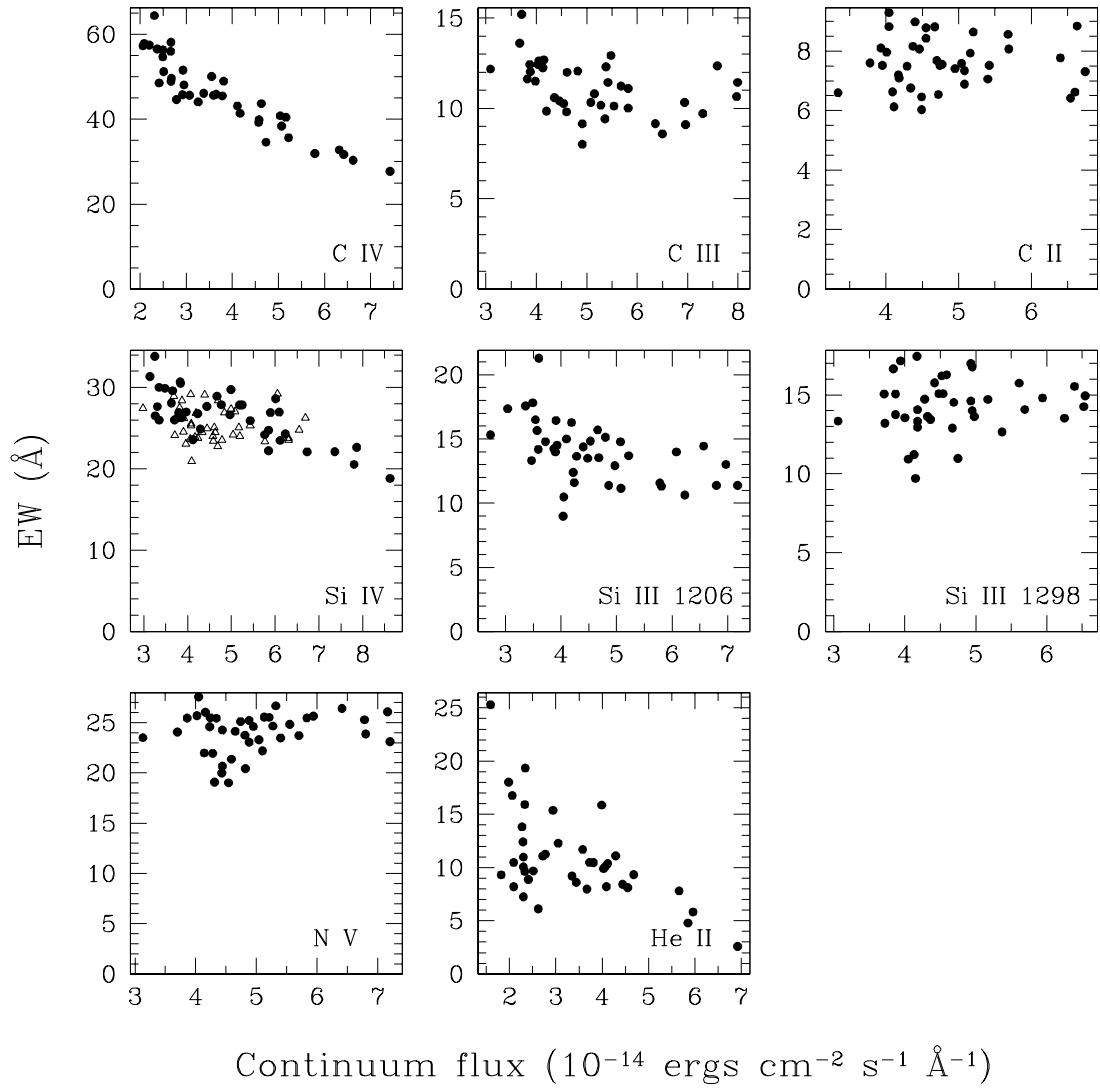}
\caption{Equivalent width versus continuum flux level for
the strong emission lines in WX~Hyi.  The open triangles and filled circles
in the \ion{Si}{4}\ plot distinguish between frames taken in the shorter
and longer wavelength grating settings, respectively. \label{fig_wxhyi_ew}}
\end{figure}

Plots of emission line fluxes and equivalent widths versus continuum fluxes
in WX~Hyi are shown in Figures~\ref{fig_wxhyi_flux} and~\ref{fig_wxhyi_ew}.
All of the emission lines except for \ion{He}{2} increase in line flux with
increased continuum level.  The continuum flux showed both short-term
flickering on the time scale of the individual spectra as well as a more
general rise in continuum level over the full observation.  The correlation
between the line and continuum fluxes is fairly tight for all of the lines,
suggesting that the line fluxes track the continuum for both the rapid
flickering and the broad continuum rise.  There is no lag between changes
in the continuum and line fluxes, although our 108~s integrations are not
sensitive to shorter time scales on which such lags might be expected to occur.

In \ion{N}{5}, \ion{Si}{3} $\lambda$1298, \ion{C}{3}, and
\ion{C}{2}, the EW remains roughly constant with the continuum
level.  The constant EWs and increasing line fluxes in these lines
indicate that their line fluxes increase equally with the
continuum flux. In \ion{C}{4}, \ion{Si}{4}, and \ion{He}{2}, the
EW of the line decreases as the continuum rises.  In \ion{C}{4}
and \ion{Si}{4}, the decline in EW is coupled with an increase in
the line flux, indicating that the lines respond to or with the
continuum but are not directly proportional to the continuum flux
level. This behavior, which has also been observed in Wolf-Rayet
star winds \citep{morris1993}, is known as the \cite{baldwin1977}
effect when observed in quasars. The decreasing EWs in \ion{C}{4}
and \ion{Si}{4} suggest that these two lines are nearly saturated.
Alternately, it could indicate a change in the ionization state of
the line formation region. \ion{N}{5} may have a higher ion
abundance than \ion{C}{4} and \ion{Si}{4} and is therefore not
near saturation, or changes in the ionization balance with
increasing flux may increase the \ion{N}{5} ion population
relative to the other two ions. The \ion{He}{2} measurements show
more scatter than for the other lines.  There is no sign of a
correlated change in the line flux with continuum, suggesting that
the \ion{He}{2} emission is decoupled from the local FUV continuum emission.

Finally, we searched for evidence of systematic changes in the
profiles of the emission lines as the continuum and line fluxes
varied.  The Gaussian fits to the emission lines showed no trend
in the line width (Gaussian FWHM) with continuum or line flux
level.  We also examined smoothed averages of the five spectra
with the lowest continuum levels and the five with the highest
continua in each grating setting.  There was no indication of a
systematic difference between the emission line shapes and no
appreciable change in the shape of the continuum between the
``low'' and ``high'' averaged spectra. The constancy of the
continuum and the emission line shapes as the flux levels increase
suggests that the variability occurs throughout the UV emission
line formation region.

\section{Discussion} \label{discussion}

\subsection{The continuum source}

The primary motivation of this study was to better constrain the
sources of FUV emission in quiescent DN.  In this respect, the
continuum modeling results have been disappointing.  WD models
provide a reasonable fit to the shape of the continuum in WX~Hyi
with realistic parameter values, but there are no actual
signatures of a WD in the spectrum.  The GHRS spectrum is of high
resolution and S/N, but the narrow absorption features present in
the model WD spectrum are not seen in the observed spectrum.  Most
of the absorption transitions are coincident with the strong
emission features and could be masked, but this is not true of
all: e.g., at 1280~\AA\ (see Figure~\ref{fig_ghrs_hyi_cont}). It
is also difficult to reconcile the prominent WD absorption lines
at \ion{C}{3} $\lambda1176$ and \ion{Si}{2} $\lambda$1262 and
$\lambda$1529 with the lack of absorption dips at zero velocity in
the relatively weak emission profiles. Furthermore, for WX~Hyi, it
would be very difficult to interpret the coordinated line and
continuum variations that are observed in the GHRS spectra in
terms of a photospheric emission from the WD.

In the case of SS~Cyg, there are some reasons for believing that
the WD should be fairly hot assuming its current behavior is
typical of the long-term average behavior.  Indeed,
\cite{schreiber2002} have argued that an implication of the {\it
HST\/}/FGS parallax of SS~Cyg is that the time averaged accretion
rate needs to be of order \expu{4.2}{17}{g~s^{-1}}, a rate that is
normally measured for nova-likes and/or the Z~Cam systems.  A
combination of compressional heating (the readjustment of the
interior structure of the WD to accreted material) and nuclear
burning of accreted material would cause the photospheric
temperature of such a system to rise to 40,000--60,000~K
\citep{townsley2003} if this is the long -term average accretion
rate.  Several Z~Cam and related systems, including MV~Lyr
\citep{hoard2004}, DW~UMa \citep{araujo2003} and Z~Cam
\citep{hartley2005} itself do have measured WD temperatures in this range.

However, as we have pointed out, to match the observed flux, model
WD spectra that fit the shape of the continuum are about 2~times
larger in WD radius than the value indicated by WD mass-radius
relationships. Another way of saying this is that a WD with the
expected radius at the known distance can only supply 25\% of the
flux at the distance.  Clearly, this result depends upon an
accurate distance to SS~Cyg, but the distance (166$_{-12}^{+14}$~pc) 
should be reliable, given that it is based upon {\it HST\/}/FGS
astrometry \citep{harrison2000}. A lower WD mass would also help
to reconcile the difference, but a WD mass of 0.68~\msun\ seems
unreasonable based on the existing optical studies
\citep{friend1990}.  Hotter WDs can match the flux but are too
blue at short wavelengths to fit the spectral shape. (To maximize
the contribution from the WD, one would look for a second source
with a lower color temperature than the WD.) As with WX~Hyi, there
are no unambiguous signatures of the WD in the FUV spectrum.
SS~Cyg has a broad Ly$\alpha$ absorption feature, but accretion
disk model fits show that Doppler-broadened absorption in (a
portion of) the disk can, in principle, explain this feature.
Again, the narrow absorption lines that appear in the normal
abundance WD models are not seen in the high S/N GHRS spectrum.

Therefore, it is evident for both SS~Cyg and for WX~Hyi that
another continuum source contributes substantially to the FUV
spectrum.  This is not unique to the emission line systems: even
in quiescent DN for which the WD dominates the FUV spectrum there
is evidence of another continuum source in the FUV.  VW~Hyi shows
a flux excess near the Lyman limit and variability, neither of
which are attributable to the WD \citep{godon2004}. For U~Gem, in
some sense the prototypical WD-dominated DN, WD model fits are
improved when a second temperature component is added --- and the
second component is required in U~Gem to reconcile flux and
temperature changes in the FUV spectrum during quiescence
\citep{long1995}. Additional absorbing material along the line of
sight to the WD is also evident in U~Gem and WZ~Sge from the
presence of transitions of \ion{S}{6}, \ion{O}{6}, \ion{N}{5}, and
\ion{Si}{4} that are not produced in their WDs and from the
improvement in model fits when an absorbing slab is placed between
the WD and the observer \citep{froning2001,long2003}.

The second source in the FUV has been variously attributed to an
``accretion belt'' on the surface of the WD, disk emission
resulting from ongoing accretion during quiescence, or a disk
corona \citep{long1993,meyer1994}. What is lacking for all of
these models are detailed calculations of expected parameter
values to compare to the observations.  The structure and physical
conditions of quiescent accretion disks remain poorly understood,
limiting our ability to select between competing models. Our
combined model fits to SS~Cyg showed that steady-state accretion
disk and simple optically thin plasma models could not provide
good fits with realistic parameter values (mass accretion rates,
temperatures, and fluxes).  The accretion disk model results are
not surprising.  Both observations and the predictions of the disk
instability model argue that quiescent disks are not in a steady
state \citep{lasota2001,horne1993}.  Plasma models are promising
but unconstrained, with the simple, single-temperatures models we
used unable to fit the shape of the spectrum. Chandra observations
of WX~Hyi show that the X-ray spectrum arises from a dense,
multi-temperature plasma, with temperatures from $10^{6}$ to
$\geq10^{8}$~K and densities of $n \sim 10^{13}$ --
$10^{14}$~cm$^{-3}$ \citep{perna2003}. Comparison of the X-ray
spectrum to standard models for quiescent DN (BL, coronas, cooling
flows) show that these models underpredict the cool gas component
and the long wavelength emission, suggesting that an
intermediate-temperature component to the quiescent emission
remains unexplored.

\subsection{The UV line emission}

The strong emission line spectra seen in WX~Hyi and SS~Cyg can be
compared to models of emission line formation in quiescent DN.
\cite{ko1996} made detailed, non-LTE calculations of the vertical
structure and emergent spectrum of a temperature-inverted layer
overlying the accretion disk in quiescent DN.  In their model, a
line-emitting accretion disk chromosphere is created by
photoionization of the disk by an external X-ray/EUV source.  In
order to generate UV line fluxes comparable to those observed in
quiescent DN, their model requires a large illuminating factor of
the disk ($\sim$0.1) and illuminating X-ray luminosities that are
as large at the outer disk radii as at the inner radii.  In
addition, they find that observed hard X-ray luminosities in DN
are too low to generate observed line fluxes using their model,
and they postulate the existence of an additional EUV/soft X-ray
component with L$_{X} \gtrsim 10^{32}$~ergs~s$^{-1}$.

In Table~\ref{tab_profiles}, we compare our observed emission line
flux ratios and radial surface brightness indices with those of
the two-component models of Ko et~al.  In their two-component
models, the hard X-ray component is fixed while the luminosity and
shape of the soft X-ray/EUV component is varied.  Given that the
models are not tailored to apply to specific DN, there is actually
fairly good agreement between the line flux ratios we observed in
SS~Cyg and WX~Hyi and those predicted by the models. The line
ratios in SS~Cyg correspond most closely with their Case~9, which
uses a 5~keV Bremsstrahlung with $L_{X}$ = \expu{9}{29}{\LUM} and
a 0.1~keV Bremsstrahlung with $L_{X} =$ \expu{1}{32}{\LUM} for the
two-component X-ray photoionizing source. The line ratios in
WX~Hyi are closer to Case~8, which has a stronger soft X-ray
component, $L_{X} =$ \expu{1}{33}{\LUM}.

The photoionization models have some deviations from the
observations, though. The models do not predict the low
\ion{N}{5}~$/$~\ion{C}{2} ratio observed in SS~Cyg, but \ion{N}{5}
is surprisingly weak in SS~Cyg compared to other DN
\citep{mauche1997}.  A more serious discrepancy is the ratio of
\ion{He}{2} to the resonance lines: the models that most closely
match the observed UV resonance line ratios in SS~Cyg and WX~Hyi
vastly over-predict the \ion{He}{2}~$/$~\ion{C}{4} ratio.  This is
a general problem with photoionization models: the flux in the
\ion{He}{2} recombination line depends directly on the strength of
the ionizing soft X-ray/EUV component and when the luminosity of
the soft component in increased to match the observed resonance
line fluxes, the \ion{He}{2} line is over-produced.  A comparison
of the radial surface brightness indices based on our fits to the
emission line profiles and those of the models also suggests that
the observed UV line surface brightnesses drop off more rapidly
with disk radius than predicted by the models.  The \cite{ko1996}
models are strictly generic, however, so models that take into
account the individual system parameters of SS~Cyg and WX~Hyi
could resolve this discrepancy.

The variability properties of the spectrum in WX~Hyi also
constrain the source of the FUV emission lines. All of the strong
emission lines, except for \ion{He}{2}, follow the variability in
the continuum with no lag on 108~sec time scales.  While most of
the lines remain constant in EW as the continuum rises, the
\ion{Si}{4} and \ion{C}{4} lines decrease in EW. These lines may
be at or nearing saturation, although it is also possible that the
ionization structure of the line-emitting region changes as the
continuum flux increases, causing Si and C to be ionized out of
these levels.  In any event, all of the resonance lines track the
FUV continuum, indicating that the two regions and formation
mechanisms are linked. The recombination line of \ion{He}{2} does
not vary with the other lines and the FUV continuum in WX~Hyi,
suggesting that variations in the soft X-ray/EUV component are not
the source of the flickering.  \cite{ko1996} note that disk
photospheric radiation is an important photoionization source for
line formation in DN, especially at small disk radii.  The radial
surface brightness profiles for both DN are steeper than the model
predictions, indicating that the observed lines are more centrally
concentrated than in the models.  Photoionization by the disk may
therefore be more significant than irradiation from the central
X-ray source for the formation of the lines in these DN, a
conclusion that is consistent with the correlation of the FUV disk
and line flux variability.

\section{Conclusions} \label{conclusions}

In this report, we have analyzed FUV spectra of SS~Cyg and WX~Hyi,
as observed with HUT and the GHRS.  The HUT spectra in particular
are the first spectra to have been obtained of these objects in
quiescence that extend to the Lyman limit.  Our results can be
summarized as follows:

\begin{enumerate}

\item The quiescent FUV spectra of WX~Hyi and SS~Cyg are
characterized by blue continua and strong, broad emission lines
from ionized species including \ion{O}{6}, \ion{N}{5}, \ion{C}{4},
\ion{Si}{4}, \ion{He}{2}, and \ion{C}{3}.  The HUT spectra of
WX~Hyi, obtained during a single interoutburst interval, show no
sign of a steady increase or decrease in FUV flux during
quiescence.

\item Model fits to the continuum using WD, steady-state accretion
disk, an optically thin plasma, and combined fits show that both
WD and disk models can fit the continuum shapes in SS~Cyg and
WX~Hyi.  However, the narrow absorption features expected from a
WD are not seen in the spectra, although the strongest transitions
are masked by line emission.  Furthermore, at the known distance
to SS~Cyg, the pure WD model fits require WD radii in excess of
that expected from WD mass-radius relations.  None of the
continuum models provide fits with realistic parameters.

\item The line profiles in SS~Cyg are double-peaked with doublet ratios
that indicate optically thin lines, while the line profiles in WX~Hyi are
single-peaked with optically thick doublet ratios.  Model fits to the line
profiles give radial surface brightness distributions of $f(R) \propto
R^{-2}$ for the FUV lines in both systems.  For SS~Cyg, the half-separation
between the emission line peaks is larger for the UV lines than for the
optical Balmer lines, indicating that the UV line formation regions is more
centrally concentrated.  Based on the velocity extent of the emission line
wings in SS~Cyg, there is no evidence for a hole in the center of the
accretion disk.

\item The spectra of SS~Cyg showed little variability during the
observations.  The GHRS spectra of WX~Hyi, on the other hand, are
highly variable in both lines and continuum on the time scale (108
sec) of the individual spectra.  There was no observed lag between
the continuum and line variations.  All of the strong lines except
\ion{He}{2} increased in flux as the continuum increased.
\ion{He}{2} showed no coordinated response to continuum
variations.  In \ion{C}{4} and \ion{Si}{4}, the EW dropped with
increasing continuum flux, providing the first examples of which
we are aware of a so-called Baldwin effect for DN, while for the
other lines the EW stayed constant. \ion{C}{4} and \ion{Si}{4} may
be near saturation, or the ionization state of UV emitting gas
changes as the continuum flux increases.  The emission line
profiles and the shape of the continuum did not change over the
observations. Since the lines presumably arise in the disk, the
coordinated variability of the lines and continuum in WX~Hyi
provides another indication of the strength of a ``second''
component in the spectrum.

\item The UV behavior of both WX~Hyi and SS~Cyg are qualitatively
in agreement with the predictions of photoionized emission line
formation models but discrepancies --- such as the weakness of
observed \ion{He}{2}\ recombination emission relative to the model
predictions --- between the observations and the models remain.

\end{enumerate}

SS Cyg (especially) and WX~Hyi are well-studied DN. Our
investigation of FUV spectra highlights our the current lack of
quantitative understanding of the physical conditions that give
rise to the FUV spectra of these two DN in quiescence.  Although
most observers (including ourselves) have tended to concentrate on
simpler WD-dominated systems in quiescence, SS~Cyg and WX~Hyi are
not unique.  Many other quiescent DNe are wholly dominated by or
have significant contributions from a ``second" component, most
probably the same component that is so prominent in SS~Cyg and 
WX~Hyi.  The exploratory work of \cite{ko1996} aside, there have been
few attempts to change the situation regarding modeling of
quiescent DN (in part because the technical challenges remain
significant). However, in the absence of such modeling, it is hard
to assert we have a good understanding of the physics of DN, and
which of several possible mechanisms give rise to the FUV emission
that is observed.  More of us need to undertake the challenge
implied by spectra of the type observed here.

\acknowledgments{We acknowledge with thanks the variable star observations
from the AAVSO International Database contributed by observers worldwide
and used in this research. This work was supported by NASA through grant
G0-6544 from the Space Telescope Science Institute, which is operated by
AURA, Inc., under NASA contract NAS5-26555.}

\clearpage
\begin{deluxetable}{clcccc}
\tablecaption{Observation Summary\label{tab_obs}}
\tablewidth{0pt}
\tablecolumns{6}
\tablehead{
\colhead{Target} & \colhead{Date (UT)} &
\colhead{Instrument} & \colhead{UT Start} &
\colhead{T$_{exp}$ (s)} & \colhead{Wavelength Range (\AA)}}
\startdata
SS~Cyg  & 1990 Dec 10 & HUT & 14:04 & \phn814\rlap{\tablenotemark{a}} & \phn827 -- 1878 \\
        & 1995 Mar 8 & HUT & 17:20 & \phn320 & \phn816 -- 1877 \\
        & 1995 Mar 8 & HUT & 17:28 & \phn768 & \phn816 -- 1877 \\
        & 1996 Sept 26 & GHRS & 14:27 & 2176 & 1149 -- 1436 \\
        & 1996 Sept 26 & GHRS & 15:55 & 2176 & 1377 -- 1664 \\
WX~Hyi  & 1995 Mar 9 & HUT & 02:34 & 1032 & \phn816 -- 1877   \\
        & 1995 Mar 10 & HUT & 13:29 & 1620 & \phn816 -- 1877  \\
        & 1995 Mar 12 & HUT & 05:15 & 1616 & \phn816 -- 1877  \\
        & 1996 Aug 5 & GHRS & 22:52 & 2176 & 1149 -- 1436 \\
        & 1996 Aug 6 & GHRS & 00:04 & 2176 & 1377 -- 1663 \\
        & 1996 Aug 6 & GHRS & 00:44 & 2176 & 1149 -- 1436 \\
        & 1996 Aug 6 & GHRS & 01:57 & 2176 & 1377 -- 1663 \\
\enddata
\tablenotetext{a}{Portion of the spectrum obtained during the night.  The
total observation time was 1476~sec.}
\end{deluxetable}

\begin{deluxetable}{lcccccccc}
\tablewidth{0pt}
\tablecaption{WD and Disk Model Fits to SS Cyg\label{sscyg_fits} }
\tablehead{
\colhead{} & \colhead{} & \colhead{} & \multicolumn{2}{c}{WD} & \colhead{}
& \multicolumn{2}{c}{Disk} \\
\cline{4-5} \cline{7-8} \\
\colhead{Model Type} &
 \colhead{Spectrum} &
 \colhead{E(B-V)} &
 \colhead{Norm\tablenotemark{a}} &
 \colhead{T$_{wd}$} &
 \colhead{} &
 \colhead{Norm\tablenotemark{b}} &
 \colhead{$log(\dot{m})$\tablenotemark{c}} &
 \colhead{$\chi^{2}/dof$}
}
\startdata
WD(norm\tablenotemark{d}) &  GHRS &  0.04 &  4.88($-$23) &  36,800 & &  \nodata &  \nodata &  6218/1277 \\
WD(DA) &  GHRS &  0.04 &  4.56($-$23) &  40,800 & &  \nodata &  \nodata &  3487/1277 \\
Disk &  GHRS &  0.04 &  \nodata &  \nodata &  & 0.163 &  16.63 &  2102/1277 \\
WD(DA)+Disk &  GHRS &  0.04 &  \nodata &  \nodata & &  0.171 &  16.66 &  2097/1277 \\
WD(norm)+Disk &  GHRS &  0.04 &  0.02($-$23) &  58,100 & & 0.174 &  16.59 &  2097/1277 \\
WD(norm) &  HUT &  0.04 &  2.69($-$23) &  48,500 & &  \nodata &  \nodata &  1342/1001 \\
WD(DA) &  HUT &  0.04 &  3.26($-$23) &  46,000 & &  \nodata &  \nodata &  1323/1001 \\
Disk &  HUT &  0.04 &  \nodata &  \nodata & &  0.075 &  17.0\phn &  1336/1001 \\
WD(DA)+Disk &  HUT &  0.04 &  3.06($-$23) &  28,700 & &  0.007 &  18.17 &  1249/1001 \\
WD(norm)+Disk &  HUT &  0.04 &  0.89($-$23) &  60,000 & &  0.136 &  16.31 &  1256/1001 \\
\enddata
\tablenotetext{a} {The WD model normalization $N=4*\pi~(R/D)^2$,
where $R_{wd}$ is the WD radius and D is the distance.}
\tablenotetext{b}{The disk models are normalized to a distance of
100 pc, and scale as $D/100$ pc$^2$} \tablenotetext{c}{$\dot{m}$ in
$g~s^{-1}$} \tablenotetext{d}{Normal abundance model.}
\end{deluxetable}

\begin{deluxetable}{lcccccccc}
\tablewidth{0pt}
\tablecaption{WD and Disk Model  Fits to WX Hyi\label{wxhyi_fits} }
\tablehead{
\colhead{} & \colhead{} & \colhead{} & \multicolumn{2}{c}{WD} & \colhead{}
& \multicolumn{2}{c}{Disk} \\
\cline{4-5} \cline{7-8} \\
\colhead{Model Type} &
 \colhead{Spectrum} &
 \colhead{E(B-V)} &
 \colhead{Norm.\tablenotemark{a}} &
 \colhead{T$_{wd}$} &
 \colhead{} &
 \colhead{Norm.\tablenotemark{b}} &
 \colhead{$log(\dot{m})$\tablenotemark{c}} &
 \colhead{$\chi^{2}/dof$}
}
\startdata
WD(norm\tablenotemark{d}) &  GHRS &  0.04 &  5.32($-$23) &  24,800 & &  \nodata &  \nodata &  3123/957 \\
WD(DA) &  GHRS &  0.04 &  1.11($-$22) &  20,700 & & \nodata &  \nodata &  1062/957 \\
Disk &  GHRS &  0.04 &  \nodata &  \nodata &  & 0.435 &  15.53 &  1286/957 \\
WD(DA)+Disk &  GHRS &  0.04 &  1.11($-$22) &  20,700 & & \nodata &  \nodata &  1062/957 \\
WD(norm)+Disk &  GHRS &  0.04 &  1.39($-$24) &  76,700 & & 1.13\phn &  15.08 &  1173/957 \\
WD(norm) &  HUT &  0.04 &  0.67($-$23) &  57,500 & & \nodata &  \nodata &  5815/989 \\
WD(DA) &  HUT &  0.04 &  0.75($-$23) &  53,800 & & \nodata &  \nodata &  5382/989 \\
Disk &  HUT &  0.04 &  \nodata &  \nodata &  & 0.011 &  16.71 &  2919/989 \\
WD(DA)+Disk &  HUT &  0.04 &  8.41($-$25) &  109,000 & & 0.141 &  15.50 &  2745/989 \\
WD(norm)+Disk &  HUT &  0.04 &  \nodata &  \nodata & & 0.06\phn &  17.02 &  5498/957 \\
\enddata
\tablenotetext{a} {The WD model normalization $N=4*\pi~(R/D)^2$,
where $R_{wd}$ is the WD radius and D is the distance.}
\tablenotetext{b}{The disk models are normalized to a distance of
100 pc, and scale as $D/100 pc^2$} \tablenotetext{c}{$\dot{m}$ in
$g~s^{-1}$} \tablenotetext{d}{Normal abundance model.}
\end{deluxetable}

\begin{deluxetable}{lccccc}
\tablewidth{0pt}
\tablecolumns{6}
\tablecaption{Observed and Predicted Line Ratios and Radial Line Profiles\label{tab_profiles}}
\tablehead{
\colhead{} & \colhead{\ion{N}{5} $\lambda$1240} &
\colhead{\ion{C}{2} $\lambda$1335} & \colhead{\ion{Si}{4} $\lambda$1400} &
\colhead{\ion{C}{4} $\lambda$1550} & \colhead{\ion{He}{2} $\lambda$1640} }
\startdata
\multicolumn{6}{c}{Line Flux Relative to \ion{C}{2}} \\
\cline{1-6} 
\noalign{\smallskip}
SS Cyg & 0.42 & 1.00 & 2.42 & 5.65 & 0.44 \\
WX Hyi & 3.32 & 1.00 & 3.19 & 4.59 & 0.86 \\
KLSK96 Case 8\tablenotemark{a} & 4.41 & 1.00 & 4.21 & 7.90 & 3.10 \\
KLSK96 Case 9\tablenotemark{a} & 1.80 & 1.00 & 2.85 & 5.27 & 2.26 \\
KLSK96 Case 10\tablenotemark{a} & 0.93 & 1.00 & 1.02 & 2.09 & 0.73 \\
\cutinhead{Surface brightness power law index $\alpha$}
SS Cyg ($\tau$ = 0)\tablenotemark{b} & \nodata & 1.77$^{+0.00}_{-0.41}$ & 2.07$^{+0.34}_{-0.12}$ & 2.35$^{+0.14}_{-0.02}$ & \nodata \\[2pt]
SS Cyg ($\tau$ = $\infty$)\tablenotemark{c} & \nodata & 1.41$^{+0.24}_{-0.09}$ & 2.03$^{+0.58}_{-0.20}$ & 2.37$^{+0.06}_{-0.16}$ & \nodata \\[2pt]
WX Hyi ($\tau$ = 0)\tablenotemark{b} & 2.34$^{+0.13}_{-0.09}$ & 1.87$^{+0.08}_{-0.10}$ & 2.22$^{+0.04}_{-0.04}$ & 2.07$^{+0.09}_{-0.12}$ & 2.07$^{+0.09}_{-0.12}$ \\[2pt]
WX Hyi ($\tau$ = $\infty$)\tablenotemark{c} & 2.71$^{+0.07}_{-0.37}$ & 1.91$^{+0.14}_{-0.10}$ & 2.48$^{+0.01}_{-0.03}$ & 1.80$^{+0.20}_{-0.10}$ & 2.08$^{+0.08}_{-0.11}$ \\
KLSK96 Case 8\tablenotemark{a} & 1.14$\pm$0.05 & 1.56$\pm$0.12 & 1.16$\pm$0.05 & 0.92$\pm$0.03 & 0.93$\pm$0.03 \\
KLSK96 Case 9\tablenotemark{a} & 1.52$\pm$0.02 & 1.85$\pm$0.09 & 1.55$\pm$0.04 & 1.22$\pm$0.04 & 1.30$\pm$0.06 \\
KLSK96 Case 10\tablenotemark{a} & 1.46$\pm$0.04 & 1.36$\pm$0.07 & 1.49$\pm$0.05 & 1.22$\pm$0.01 & 1.30$\pm$0.07 \\
\cutinhead{$R_{outer} / R_{inner}$ of line emitting region}
SS Cyg ($\tau$ = 0)\tablenotemark{b} & \nodata & 17.2$^{+82.8}_{-0.67}$ & 24.8$^{+29.3}_{-7.90}$ & 35.9$^{+19.3}_{-6.7}$ & \nodata \\[2pt]
SS Cyg ($\tau$ = $\infty$)\tablenotemark{c} & \nodata & 100.0$^{+0.0}_{-81.2}$ & 21.3$^{+29.2}_{-4.8}$ & 33.4$^{+7.7}_{-8.2}$ & \nodata \\[2pt]
WX Hyi ($\tau$ = 0)\tablenotemark{b} & 98.6$^{+1.4}_{-90.4}$ & 37.6$^{+23.1}_{-7.0}$ & 88.0$^{+12.0}_{-46.8}$ & 43.1$^{+56.9}_{-4.8}$ & 100.0$^{+0.0}_{-0.02}$ \\[2pt]
WX Hyi ($\tau$ = $\infty$)\tablenotemark{c} & 24.8$^{+75.2}_{-17.6}$ & 40.2$^{+23.4}_{-6.2}$ & 100.00$^{+0.0}_{-31.5}$ & 73.2$^{+0.0}_{-33.2}$ & 100.0$^{+0.01}_{-12.6}$ \\
\cutinhead{$V_{disk} \sin \imath$ (km~s$^{-1}$)}
SS Cyg ($\tau$ = 0)\tablenotemark{b} & \nodata &  490$^{+75}_{-33}$ & 470$^{+52}_{-44}$ & 280$^{+19}_{-60}$ & \nodata \\[2pt]
SS Cyg ($\tau$ = $\infty$)\tablenotemark{c} & \nodata & 520$^{+34}_{-90}$ & 470$^{+69}_{-230}$ & 280$^{+62}_{-30}$ & \nodata \\[2pt]
WX Hyi ($\tau$ = 0)\tablenotemark{b} & 210$^{+350}_{-25}$ & 380$^{+34}_{-37}$ & 310$^{+53}_{-24}$ & 340$^{+3}_{-180}$ & 330$^{+73}_{-40}$ \\[2pt]
WX Hyi ($\tau$ = $\infty$)\tablenotemark{c} & 310$^{+290}_{-160}$ & 360$^{+36}_{-52}$ & 160$^{+34}_{-5.1}$ & 350$^{+44}_{-33}$ & 310$^{+61}_{-43}$ \\
\enddata
\tablenotetext{a}{~Line flux and radial line emissivity predictions from the models of \cite{ko1996}, labelled here for clarity as KLSK96. Cases~8~-- 10 invoke a two-component model for the photoionizing continuum. The hard component in all three cases is a 5~keV Bremsstrahlung with L$_{X}$ = 9$\times10^{29}$ ergs~s$^{-1}$.  For Case~8, the soft component is a 0.1~keV Bremsstrahlung with L$_{X}$ = 10$^{33}$ ergs~s$^{-1}$. For Case~9, the soft component is a 0.1~keV Bremsstrahlung with L$_{X}$ = 10$^{32}$ ergs~s$^{-1}$.  For Case~10, the soft component is a 5$\times10^{5}$ blackbody with L$_{X}$ = 10$^{32}$ ergs~s$^{-1}$. The UV line flux ratios have been taken from Table~7, while the radial power law emissivity indices have been derived from the UV line fluxes in Table~4 of \cite{ko1996}.} \tablenotetext{b}{Power law indices for models marked $\tau = 0$ have been derived from the observations under the assumption that the lines are optically thin.} \tablenotetext{c}{Power law indices for models marked
$\tau = \infty$ have been derived from the observations under the assumption that the lines are optically thick.}
\end{deluxetable}

\end{document}